\def\RecRSIShowAuthors{1} 
\PassOptionsToClass{balance=false}{acmart}
\ifdefined\RecRSIShowAuthors
  \documentclass[sigconf]{acmart}
\else
  \documentclass[sigconf,anonymous,review]{acmart}
\fi

\usepackage{algorithm}
\usepackage{algpseudocode}
\usepackage{placeins}
\usepackage{needspace}
\usepackage{tikz}
\usepackage{multirow}
\usepackage{stfloats}
\usepackage{cuted}
\usepackage{capt-of}
\setcopyright{none}
\renewcommand\footnotetextcopyrightpermission[1]{}
\acmConference{}{}{}
\acmBooktitle{}
\acmDOI{}
\acmISBN{}

\AtBeginDocument{\fancyhead[LE]{}\fancyhead[RO]{}}

\begin{document}
\hypersetup{pdftitle={RecRSI: Mechanism-Guided Discovery of Recommendation Algorithms}}
\raggedbottom
\ifdefined\RecRSIShowAuthors
\else
  \makeatletter
  \fancyhead[LE]{\ACM@linecountL}
  \fancyhead[RO]{\ACM@linecountR}
  \makeatother
\fi
\title{RecRSI: Mechanism-Guided Discovery of Recommendation Algorithms}
\ifdefined\RecRSIShowAuthors

\author{Tingran Gan}
\authornote{Both authors contributed equally to this research.}
\email{dynamisg@sjtu.edu.cn}
\affiliation{%
  \institution{Shanghai Jiao Tong University}
  \city{Shanghai}
  \country{China}
}

\author{Zhouziyue Lu}
\authornotemark[1]
\email{zhouziyue\_lu@sjtu.edu.cn}
\affiliation{%
  \institution{Shanghai Jiao Tong University}
  \city{Shanghai}
  \country{China}
}

\author{Dian Jin}
\email{2430XH10019@m.fudan.edu.cn}
\affiliation{%
  \institution{Fudan University}
  \city{Shanghai}
  \country{China}
}

\author{Tao Li}
\email{litao118@meituan.com}
\affiliation{%
  \institution{Meituan}
  \city{Beijing}
  \country{China}
}

\author{Yang Gao}
\email{gaoyang94@meituan.com}
\affiliation{%
  \institution{Meituan}
  \city{Beijing}
  \country{China}
}

\author{Ming Liu}
\email{liuming66@meituan.com}
\affiliation{%
  \institution{Meituan}
  \city{Beijing}
  \country{China}
}

\author{Weiwen Liu}
\authornote{Corresponding authors.}
\email{wwliu@sjtu.edu.cn}
\affiliation{%
  \institution{Shanghai Jiao Tong University}
  \city{Shanghai}
  \country{China}
}

\author{Jianghao Lin}
\authornotemark[2]
\email{linjianghao@sjtu.edu.cn}
\affiliation{%
  \institution{Shanghai Jiao Tong University}
  \city{Shanghai}
  \country{China}
}

\author{Weinan Zhang}
\authornotemark[2]
\email{wnzhang@sjtu.edu.cn}
\affiliation{%
  \institution{Shanghai Jiao Tong University}
  \city{Shanghai}
  \country{China}
}

\author{Yong Yu}
\authornotemark[2]
\email{yyu@sjtu.edu.cn}
\affiliation{%
  \institution{Shanghai Jiao Tong University}
  \city{Shanghai}
  \country{China}
}

\renewcommand{\shortauthors}{Gan et al.}

\fi

\begin{abstract}
Improving recommendation algorithms for the Web requires coordinating interacting design choices and learning from experiments.
We present RecRSI, an LLM-based research framework that uses mechanism hypotheses to organize algorithm discovery.
Each hypothesis links a proposed mechanism change to a reference design, a rationale, and an expected observation.
MechSpace derives implementable primitives and composition rules from the task source; Research Line uses this space and prior evidence to develop and test hypotheses; Mechanism Feedback carries outcomes and their design context into subsequent research.
This organization makes each experiment inform which mechanisms to develop, revise, or combine next.
Across four initial recommenders and four datasets, RecRSI yields average relative improvements of 8.48\% in NDCG@5 and 6.88\% in HR@5 over the initial algorithms, and 2.60\% and 2.20\%, respectively, over the strongest competing result for each setting and metric.
Component and process studies link these gains to coordinated mechanism choices and evidence-guided follow-up.
Our code is publicly available at \url{https://github.com/DynamisG210/RecRSI}.

\end{abstract}



\maketitle

\providecommand{\RecRSIFigure}[1]{%
  \IfFileExists{#1}{\includegraphics[width=\textwidth]{#1}}{%
    \PackageWarning{RecRSI}{Current figure #1 is missing; use the figure from the delivered source package}%
    \fbox{\parbox[c][35mm][c]{0.96\textwidth}{\centering
      Current author-provided vector figure\\[4pt]
      \texttt{\detokenize{#1}}\\[4pt]
      The complete source package includes this asset.}}%
  }%
}

\section{Introduction}
\label{sec:introduction}

Recommendation algorithms shape how users discover content and products across the Web, from personalized feeds to online marketplaces.
Developing these ranking algorithms across different interaction settings requires coordinating representations, scoring functions, learning objectives, and optimization procedures, whose effectiveness depends on the task and on one another~\cite{mao2021simplex}.
With limited training and evaluation budgets, automated research must identify worthwhile changes and use their outcomes to guide further discovery.

Large language model (LLM) agents increasingly automate algorithm design and experimentation.
Program-evolution systems search over evaluated candidates~\cite{novikov2025alphaevolve,lange2025shinkaevolve}, while research agents develop and test hypotheses~\cite{lu2024aiscientist,yamada2025aiscientistv2,jiang2025aide}.
In recommendation, Self-EvolveRec couples user simulation with model diagnosis, while RecHarness combines direction routing with accumulated experimental skills~\cite{kim2026selfevolverec,ling2026recharness}.
Sustained algorithm discovery poses two challenges for these systems.

\begin{figure*}[t]
  \centering
  \RecRSIFigure{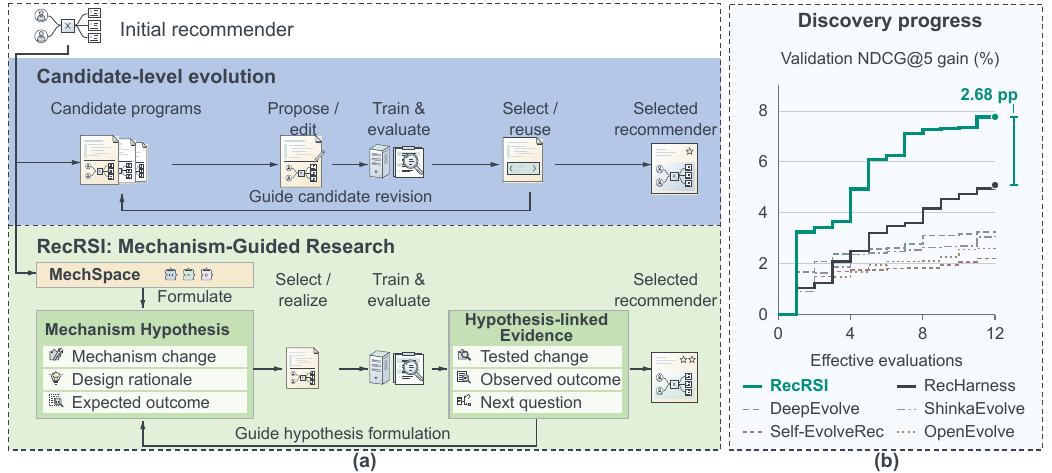}
  \caption{Research organization and algorithm discovery. (a) Candidate-level evolution organizes iteration around programs; RecRSI uses mechanism hypotheses to connect proposed changes, their rationale, and experimental evidence for the next design. Both workflows generate, evaluate, and refine algorithms. (b) Mean best-so-far validation NDCG@5 gain over the initial recommender on the same eight tasks as Figure~\ref{fig:discovery-cost}(a,b). Search repeats are averaged within each task, then tasks receive equal weight.}
  \Description{The left panel contrasts candidate-level evolution with RecRSI. MechSpace supplies interventions for a mechanism hypothesis, which connects a change, rationale, and expected outcome. Evaluation produces hypothesis-linked evidence that guides the next hypothesis. The right panel compares eight-task algorithm discovery progress for RecRSI and five systems, using the mean relative gain in best-so-far validation NDCG at rank five. Repeats are averaged within tasks before equally weighting tasks.}
  \label{fig:intro}
\end{figure*}

\textbf{Interdependent mechanisms obscure which changes to test.}
Studies of recommendation architectures and interaction search show that useful component choices depend on their task and surrounding design~\cite{he2020lightgcn,rendle2020revisited,song2020autoctr}.
A scoring change, for example, affects both prediction and the scores consumed by the training objective.
Research must decide which mechanisms to change together and how to implement their interaction.
Distinct code edits may revisit similar mechanisms; a promising combination may span several components.
Repeatedly reconstructing these relationships during program search makes targeted experimentation difficult within a limited budget.

\textbf{Context-dependent outcomes are difficult to reuse.}
Reflection-based agents reuse feedback across trials~\cite{shinn2023reflexion,ye2024reevo}, and CausalEvolve emphasizes organizing past experience to guide further evolution~\cite{chen2026causalevolve}.
In recommendation, an evaluation measures a complete algorithm under a particular mechanism combination.
Its value for the next experiment depends on the tested hypothesis, reference design, and unchanged components.
A failed alternative may challenge one explanation while leaving an earlier design worth pursuing.
Research must connect the observation to that design context to decide what to refine, replace, or recombine.

A \emph{mechanism} is a design with a specific computational or learning role, such as sequence aggregation, matching geometry, or a ranking objective.
A \emph{mechanism hypothesis} links a proposed change to a reference design, a rationale, and an expected observation.
It makes an experiment both a concrete algorithmic intervention and a research question whose outcome can inform the next design.

We introduce \textbf{RecRSI}, a framework that uses mechanism hypotheses as the organizing unit of recommendation algorithm discovery.
\textbf{MechSpace} derives task-specific primitives and composition rules, making proposed interventions implementable.
\textbf{Research Line} develops and tests hypotheses through complementary roles for composition, refinement, explanation testing, and exploration.
\textbf{Mechanism Feedback} interprets outcomes in the context of the tested hypotheses and designs to inform the next research decision.
A new hypothesis can build on a successful earlier design and use an unsuccessful experiment to choose its next intervention.
Section~\ref{sec:method_feedback} demonstrates this process in a real research run.
Figure~\ref{fig:intro} links this hypothesis-centered organization to algorithm discovery progress across eight tasks.
The name reflects a recursive self-improvement perspective~\cite{zhang2025dgm}: evaluated recommender designs and their evidence support successive algorithm improvements.

Experiments cover four initial recommenders and four datasets.
Across the 16 model--dataset settings, RecRSI improves NDCG@5 and HR@5 by an average of 8.48\% and 6.88\% over the initial algorithms, and 2.60\% and 2.20\% over the strongest competing result for each setting and metric.
Component and process studies trace these gains to coordinated mechanism choices, history-informed follow-up, and discoveries beyond the initial search horizon.
Our contributions are:
\begin{itemize}
  \item \textbf{A mechanism-guided discovery framework.} RecRSI makes mechanism hypotheses the research unit connecting algorithm design, experimentation, and follow-up.
  \item \textbf{A method for cumulative mechanism research.} Task-derived interventions and hypothesis-linked feedback turn earlier successes and failures into concrete follow-up experiments.
  \item \textbf{Evidence of discovery quality and research value.} Evaluation across 16 settings establishes ranking gains; component, history, and continuation studies connect them to the research decisions and mechanism combinations that produce them.
\end{itemize}

\section{Related Work}
\label{sec:related_work}

\subsection{Automated Algorithm Discovery and Program Evolution}
Automatic algorithm design searches over executable computations, from AutoML-Zero's mathematical operations to FunSearch's LLM-generated programs~\cite{real2020automlzero,romeraparedes2024funsearch}.
Evolution of Heuristics (EoH) co-evolves ideas and code~\cite{liu2024eoh}; AlphaEvolve, OpenEvolve, and ShinkaEvolve develop populations of evaluated programs with complementary selection and diversity mechanisms~\cite{novikov2025alphaevolve,sharma2025openevolve,lange2025shinkaevolve}.
The Darwin G\"odel Machine extends this loop to the coding agent's own implementation~\cite{zhang2025dgm}.

Self-Refine iterates on generated outputs, Reflexion stores verbal feedback for subsequent trials, and ReEvo uses short- and long-term reflections in evolution~\cite{madaan2023selfrefine,shinn2023reflexion,ye2024reevo}.
DeepEvolve incorporates literature-informed hypotheses; CausalEvolve organizes outcome- and procedure-level factors in a causal scratchpad and uses unexpected outcomes to generate follow-up hypotheses~\cite{liu2025deepevolve,chen2026causalevolve}.
The AI Scientist automates research from ideas through experiments and manuscripts; its successor and AIDE organize experimentation through tree search~\cite{lu2024aiscientist,yamada2025aiscientistv2,jiang2025aide}.
The autoresearch software provides a compact code--train--evaluate loop with a fixed training-time budget~\cite{karpathy2026autoresearch}.
Broader reviews examine memory, skills, protocols, and harnesses as externalized agent capabilities~\cite{zhou2026externalization}, and analyze agentic data generation through accuracy, complexity, and diversity~\cite{zeng2026agenticdata}.

AgentNAS builds interchangeable module slots; FairNAD structures literature-derived design attributes~\cite{jeong2026agentnas,sakuma2026fairnad}.
MechSpace derives task-specific mechanism specifications and composition rules, connecting a proposed research question to an implementable intervention.

\subsection{Automated Research for Recommender Systems}
RecBole provides unified implementations and protocols~\cite{zhao2021recbole}; systematic studies underline the importance of tuning and comparable evaluations~\cite{dacrema2019progress,rendle2020revisited}.
AutoFIS selects feature interactions, and AutoCTR searches their architectural combinations~\cite{liu2020autofis,song2020autoctr}.

LLMs support multiple stages of recommendation, from feature modeling and ranking to pipeline control~\cite{lin2025llmrecsurvey}.
Self-EvolveRec couples simulated-user critiques with model diagnostics~\cite{kim2026selfevolverec}.
RecHarness combines bandit routing with accumulated Experiment Skills, grouped trials, validation-based reward updates, and structural jumps followed by local retuning~\cite{ling2026recharness}.
EvoRec distills methodologies from experiment memory; AgentX-Model coordinates reproduction, follow-up, composition, and diagnosis~\cite{mu2026evorec,yang2026agentxmodel}.
RecRSI organizes recommendation research around a mechanism hypothesis specifying an implementable intervention, its reference design, a rationale, and an expected observation.
This hypothesis guides candidate construction and links the resulting evidence to the mechanisms changed and held fixed, informing which designs to refine or recombine next.

\providecommand{\RecRSIFigure}[1]{%
  \IfFileExists{#1}{\includegraphics[width=\textwidth]{#1}}{%
    \PackageWarning{RecRSI}{Current figure #1 is missing; use the figure from the delivered source package}%
    \fbox{\parbox[c][35mm][c]{0.96\textwidth}{\centering
      Current author-provided vector figure\\[4pt]
      \texttt{\detokenize{#1}}\\[4pt]
      The complete source package includes this asset.}}%
  }%
}

\section{Methodology}
\label{sec:methodology}

\begin{figure*}[!t]
  \centering
  \RecRSIFigure{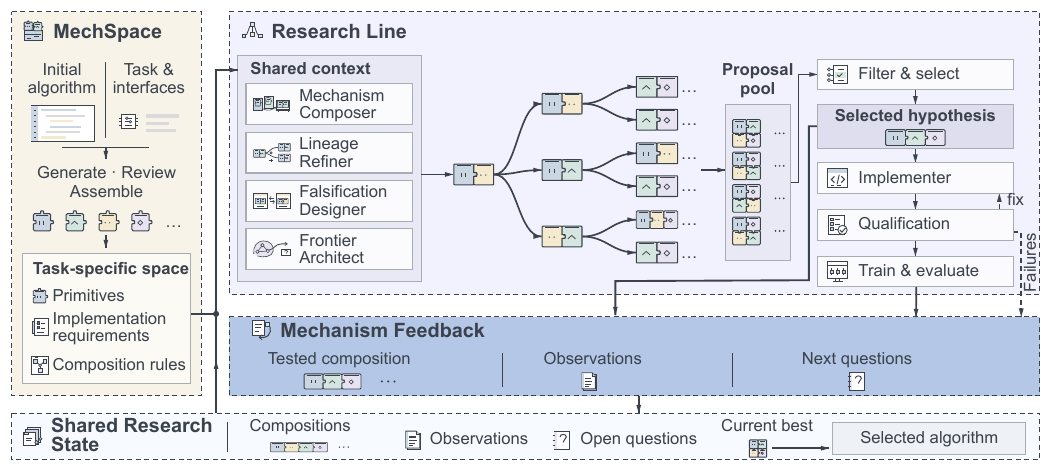}
  \caption{Overview of RecRSI. MechSpace supplies the design space for mechanism hypotheses; Research Line develops them into candidates and experiments; Mechanism Feedback links outcomes to the tested hypotheses and designs. The shared research state carries this evidence into subsequent investigations.}
  \Description{Task source and interfaces enter MechSpace. Four research roles develop mechanism hypotheses for selection, implementation, and evaluation. Feedback links observations to the tested hypotheses and reference designs. Research records, open questions, and the current best form the shared state used in subsequent rounds.}
  \label{fig:method}
\end{figure*}

RecRSI carries a mechanism hypothesis through formulation, implementation, evaluation, and follow-up.
Together, its three stages connect a research question to an executable intervention and turn the outcome into evidence for the next decision (Figure~\ref{fig:method}).

\subsection{Task and Mechanism Hypotheses}
\label{sec:method_task}

Given an initial recommendation algorithm $A_0$, a task environment $\mathcal T$, and a budget $B$, RecRSI searches for an improved algorithm implementation.
Researchers supply the initial source, data splits, editable interfaces, and a fixed training--evaluation protocol $\Pi$:
\begin{equation}
\theta_A=\operatorname{Train}_{\Pi}(A;D_{\mathrm{tr}}),
\qquad
q(A)=\operatorname{Eval}_{\Pi}(A,\theta_A;D_{\mathrm{val}}).
\label{eq:method_quality}
\end{equation}
The inner process learns parameters; the outer process changes the algorithm under limits on experimental opportunities and resources.
RecRSI returns the highest-scoring evaluated algorithm, including $A_0$ among the candidates.

A composition $C$ describes the mechanisms used together and their relationships.
A mechanism hypothesis is $h_t=(C_t^{\mathrm{ref}},\allowbreak C_t,\allowbreak \rho_t,\allowbreak \epsilon_t)$: the reference design, proposed composition, research rationale, and expected observation.
The transition from $C_t^{\mathrm{ref}}$ to $C_t$ specifies which mechanisms stay unchanged and which are added, replaced, or removed; $\rho_t$ and $\epsilon_t$ state why this intervention merits an experiment and what it is expected to reveal.
The hypothesis guides implementation of candidate $A_t$; $\widehat C_t$ denotes the composition described in the implemented candidate's record.

The research state $\mathcal R_t=(\mathcal H_t,\mathcal U_t,A_t^\star)$ contains hypothesis--experiment records, open questions, and the current best algorithm.
Records store both the construction references and the observations motivating follow-up hypotheses.

\subsection{MechSpace: A Design Space for Mechanism Hypotheses}
\label{sec:method_mechspace}

MechSpace converts the initial source and task interfaces into a mechanism space $\mathcal S_{\mathcal T}=(\mathcal P_{\mathcal T},\Gamma_{\mathcal T})$ of primitives and composition rules.
Each primitive connects a computational or learning effect to a source location, input and output types, parameters, and implementation requirements.
Its effect path specifies how the changed computation participates in training or prediction.

Within the supplied edit scope and composition limits, an LLM generator proposes primitives, parameter settings, source anchors, and compatibility rules.
An LLM reviewer examines coverage, redundancy, composability, and effect paths from the supplied source; a deterministic assembler validates schemas, mutation targets, and composition constraints to produce the registry and grammar.
Effect-path assessment is source-based LLM review; candidate execution is checked separately during qualification.
For example, the reviewer rejects an evaluation-only learned readout because the training loss does not use its new parameters.
The MoRec/CDs space covers representation, scoring, objective, update, architecture, and sampling; it is built once and reused through task adapters.

\paragraph{Composing interventions.}
The grammar determines how primitives can be combined.
In the MoRec/CDs instance, \texttt{SELECT} chooses one primitive, and \texttt{COMPOSE} combines two or three from distinct axes, subject to mutation-target constraints.
The primitive \texttt{score\_\allowbreak cosine\_\allowbreak temp} changes \texttt{Model.\allowbreak predict\_\allowbreak score}, shared by training and candidate evaluation.
The primitive \texttt{obj\_\allowbreak pairwise\_\allowbreak margin} adds a ranking term through the same positive and negative scores:
\begin{equation}
\begin{aligned}
s(u,i)&=\langle\bar{\mathbf z}_u,\bar{\mathbf e}_i\rangle/\tau,\\
\mathcal L&=\mathcal L_{\mathrm{BCE}}+
\frac{\lambda}{|\mathcal V|}\sum_{j\in\mathcal V}
\operatorname{softplus}\!\left(m-(s_j^+-s_j^-)\right).
\end{aligned}
\label{eq:mechanism_example}
\end{equation}
Here $\mathbf z_u$ and $\mathbf e_i$ are sequence and item representations, with $\bar{\mathbf v}=\mathbf v/(\|\mathbf v\|_2+\varepsilon)$.
$\mathcal V$ indexes non-padding supervised positions, $s_j^+$ and $s_j^-$ are their positive and negative scores, and $\mathcal L_{\mathrm{BCE}}$ sums the mean positive and negative BCE losses.
This instance uses $\tau=m=0.2$, $\lambda=0.5$, and $\varepsilon=10^{-8}$.
Sharing the scores makes the interaction between scoring and learning explicit.
Focal weighting offers an alternative on the objective axis, while recent-history fusion modifies representations upstream of the scorer.

\begin{algorithm}[!t]
\caption{Mechanism-guided recommendation research}
\label{alg:recrsi}
\begin{algorithmic}[1]
\Require Initial algorithm $A_0$, task $\mathcal T$, budget $B$
\State $\mathcal S \gets \operatorname{MechSpace}(A_0,\mathcal T)$
\State $\mathcal R \gets \operatorname{Initialize}(A_0,\mathcal T)$
\While{research budget remains}
  \State $\mathcal Q \gets \operatorname{Propose}(\mathcal S,\mathcal R,B)$
  \State $(h,C) \gets \operatorname{SelectFeasible}(\mathcal Q,\mathcal R,B)$
  \State \textbf{if} no eligible proposal \textbf{then break}
  \State $(A,\widehat C,z) \gets \operatorname{ImplementAndQualify}(h,C,\mathcal T)$
  \If{$z=\mathrm{PASS}$}
    \State $o \gets \operatorname{AdmitAndEvaluate}_{\Pi}(A,B)$
  \Else
    \State $o \gets \operatorname{QualificationFailure}(z)$ \Comment{skip evaluation}
  \EndIf
  \State $e \gets \operatorname{Feedback}(\mathcal T,h,\widehat C,o;\mathcal R)$
  \State Update $\mathcal R$, current best, and remaining $B$
\EndWhile
\State \Return Current best algorithm in $\mathcal R$
\end{algorithmic}
\end{algorithm}

\subsection{Research Line: Advancing Mechanism Hypotheses}
\label{sec:method_research_line}

Four complementary roles form a hypothesis pool $\mathcal Q$ from the mechanism space and research state (Algorithm~\ref{alg:recrsi}).
The \emph{Mechanism Composer} proposes combinations with a joint rationale; the \emph{Lineage Refiner} develops an existing direction; the \emph{Falsification Designer} formulates contrasts between explanations; and the \emph{Frontier Architect} identifies underexplored directions while considering its peers' proposals.

\paragraph{From a hypothesis to a complete design.}
Selected primitives specify the intervention; the reference candidate specifies the surrounding design.
A scoring hypothesis, for example, also identifies the objective to use or restore when revisiting an earlier branch.
The task adapter resolves this complete composition against the mechanism space and editable interfaces.
Feasibility checks enforce composition rules and reference links; candidate identities support duplicate filtering.

The roles adapt their hypotheses to the available evidence; a fixed selector allocates experimental opportunities.
Within each opportunity block, it enforces role coverage at the opening positions and whenever the remaining positions are needed to cover available, unselected roles.
Other choices follow a fixed role-priority order with stable tie-breaking, reserving opportunities for complementary research actions.

The Implementer receives the hypothesis specification, source, and interface requirements.
Qualification checks interfaces and execution, with bounded repair for implementation failures.
Candidates that pass qualification and resource admission enter training and evaluation under $\Pi$.
An unresolved qualification failure skips evaluation and records its status, diagnostics, and cost; the incumbent stays unchanged and research proceeds to the next eligible proposal under the remaining budget.
These records and measured outcomes feed Mechanism Feedback.

\begin{table*}[!t]
\centering
\caption{Final test performance across four seed recommenders and four datasets.}
\label{tab:overall}
\small
\renewcommand{\arraystretch}{0.88}
\setlength{\tabcolsep}{4pt}
\resizebox{\textwidth}{!}{%
\begin{tabular}{llcccccccc}
\toprule
\multirow{2}{*}{Seed Recommender} & \multirow{2}{*}{Method} & \multicolumn{2}{c}{CDs} & \multicolumn{2}{c}{Electronics} & \multicolumn{2}{c}{Office} & \multicolumn{2}{c}{MovieLens-1M} \\
\cmidrule(lr){3-4} \cmidrule(lr){5-6} \cmidrule(lr){7-8} \cmidrule(lr){9-10}
 &  & NDCG@5 & HR@5 & NDCG@5 & HR@5 & NDCG@5 & HR@5 & NDCG@5 & HR@5 \\
\midrule
\multirow{7}{*}{SASRec} & Seed & 0.3489 & 0.4572 & 0.3623 & 0.4860 & 0.2674 & 0.3980 & 0.5613 & 0.7176 \\
 & RecHarness & \underline{0.3954} & \underline{0.5126} & \underline{0.3705} & \underline{0.4947} & 0.2754 & 0.4080 & \textbf{0.5687} & \underline{0.7241} \\
 & DeepEvolve & 0.3510 & 0.4605 & 0.3678 & 0.4932 & \underline{0.2795} & 0.4099 & 0.5340 & 0.6917 \\
 & OpenEvolve & 0.3645 & 0.4841 & 0.3674 & 0.4920 & 0.2771 & \textbf{0.4139} & 0.5610 & 0.7188 \\
 & ShinkaEvolve & 0.3641 & 0.4841 & 0.3629 & 0.4871 & 0.2743 & 0.4100 & 0.5647 & 0.7212 \\
 & Self-EvolveRec & 0.3645 & 0.4806 & 0.3679 & 0.4918 & 0.2731 & 0.4055 & 0.5607 & 0.7171 \\
 & RecRSI (Ours) & \textbf{0.4329} & \textbf{0.5605} & \textbf{0.3730} & \textbf{0.5027} & \textbf{0.2811} & \underline{0.4127} & \underline{0.5678} & \textbf{0.7275} \\
 & \textit{Improv. (\%)} & $+9.50$ & $+9.36$ & $+0.67$ & $+1.61$ & $+0.58$ & $-0.28$ & $-0.16$ & $+0.47$ \\
\midrule
\multirow{7}{*}{MoRec} & Seed & 0.3795 & 0.5014 & 0.3698 & 0.4949 & 0.2887 & 0.4258 & 0.5636 & 0.7209 \\
 & RecHarness & \underline{0.4011} & \underline{0.5283} & 0.3726 & 0.4982 & 0.2855 & 0.4240 & 0.5687 & 0.7264 \\
 & DeepEvolve & 0.3863 & 0.5104 & \underline{0.3804} & \underline{0.5053} & 0.2875 & 0.4259 & \textbf{0.5715} & \underline{0.7267} \\
 & OpenEvolve & 0.3924 & 0.5203 & 0.3780 & 0.5033 & \textbf{0.2905} & \textbf{0.4291} & 0.5651 & 0.7234 \\
 & ShinkaEvolve & 0.3875 & 0.5133 & 0.3774 & 0.5026 & 0.2897 & 0.4270 & 0.5671 & 0.7240 \\
 & Self-EvolveRec & 0.3878 & 0.5134 & 0.3730 & 0.4982 & 0.2901 & \underline{0.4277} & 0.5645 & 0.7212 \\
 & RecRSI (Ours) & \textbf{0.4151} & \textbf{0.5416} & \textbf{0.3938} & \textbf{0.5231} & \underline{0.2902} & 0.4267 & \underline{0.5708} & \textbf{0.7271} \\
 & \textit{Improv. (\%)} & $+3.50$ & $+2.52$ & $+3.51$ & $+3.52$ & $-0.11$ & $-0.56$ & $-0.12$ & $+0.05$ \\
\midrule
\multirow{7}{*}{BERT4Rec} & Seed & 0.3286 & 0.4417 & 0.3881 & 0.5072 & 0.2423 & 0.3643 & 0.5313 & 0.6809 \\
 & RecHarness & 0.3220 & 0.4383 & \underline{0.3959} & \underline{0.5148} & 0.2363 & 0.3572 & \textbf{0.5476} & \textbf{0.6954} \\
 & DeepEvolve & \underline{0.3334} & \underline{0.4480} & 0.3905 & 0.5091 & 0.2374 & 0.3574 & 0.5403 & 0.6893 \\
 & OpenEvolve & 0.3286 & 0.4417 & 0.3881 & 0.5072 & 0.2423 & 0.3643 & 0.5344 & 0.6833 \\
 & ShinkaEvolve & 0.3281 & 0.4418 & 0.3932 & 0.5123 & \underline{0.2451} & \underline{0.3678} & 0.5425 & 0.6882 \\
 & Self-EvolveRec & 0.3070 & 0.4146 & 0.3807 & 0.4984 & 0.2169 & 0.3280 & \underline{0.5464} & \underline{0.6937} \\
 & RecRSI (Ours) & \textbf{0.3742} & \textbf{0.4940} & \textbf{0.4046} & \textbf{0.5233} & \textbf{0.2681} & \textbf{0.3963} & 0.5424 & 0.6905 \\
 & \textit{Improv. (\%)} & $+12.24$ & $+10.27$ & $+2.20$ & $+1.66$ & $+9.36$ & $+7.76$ & $-0.96$ & $-0.70$ \\
\midrule
\multirow{7}{*}{NGCF} & Seed & 0.2919 & 0.4059 & 0.2642 & 0.3635 & 0.0867 & 0.1365 & 0.2771 & 0.6547 \\
 & RecHarness & 0.2916 & 0.4028 & \textbf{0.2996} & \textbf{0.4064} & 0.0916 & 0.1430 & 0.2778 & 0.6606 \\
 & DeepEvolve & 0.3217 & 0.4380 & 0.2875 & 0.3899 & 0.0959 & 0.1454 & 0.2883 & 0.6727 \\
 & OpenEvolve & \underline{0.3353} & \underline{0.4503} & 0.2890 & 0.3914 & 0.1011 & 0.1519 & \underline{0.2939} & 0.6838 \\
 & ShinkaEvolve & 0.3349 & 0.4497 & 0.2873 & \underline{0.3946} & \underline{0.1022} & \underline{0.1531} & \underline{0.2939} & \underline{0.6850} \\
 & Self-EvolveRec & 0.3189 & 0.4343 & 0.2725 & 0.3738 & 0.0950 & 0.1446 & 0.2871 & 0.6682 \\
 & RecRSI (Ours) & \textbf{0.3366} & \textbf{0.4514} & \underline{0.2922} & 0.3936 & \textbf{0.1051} & \textbf{0.1566} & \textbf{0.2957} & \textbf{0.6863} \\
 & \textit{Improv. (\%)} & $+0.41$ & $+0.24$ & $-2.45$ & $-3.15$ & $+2.86$ & $+2.23$ & $+0.63$ & $+0.20$ \\
\bottomrule
\end{tabular}%
}
\vspace{2pt}
\begin{minipage}{\textwidth}\footnotesize
Bold/underline: best/second best. Improv.: RecRSI's relative gain over the strongest competing system, computed before rounding.
\end{minipage}
\end{table*}

\subsection{Mechanism Feedback and Research State Update}
\label{sec:method_feedback}

Mechanism Feedback relates an outcome to the intervention and question that produced it.
Each record links the hypothesis $h_t$, implemented composition $\widehat C_t$, and observation $o_t$ to the reference design, changed and unchanged mechanisms, performance difference, execution status, and cost.
The original rationale, expected observation, and suggested tests remain available alongside the measurements.

\paragraph{Using evidence across experiments.}
The feedback record updates the shared state:
\begin{equation}
\begin{aligned}
e_t &= \operatorname{Feedback}(\mathcal T,h_t,\widehat C_t,o_t;\mathcal R_t),\\
\mathcal R_{t+1} &= \operatorname{Update}(\mathcal R_t,e_t).
\end{aligned}
\label{eq:method_feedback}
\end{equation}
Outcomes update the experimental history and mechanism-level evidence; the current best changes when validation performance improves.
The construction reference determines the base design; evidence from another experiment can motivate a new intervention.
Role-specific inputs include the incumbent, latest result, recent mechanism comparisons, open questions, and role history, linking working explanations to measured outcomes.

\paragraph{An unsuccessful hypothesis guides the next discovery.}
In a MoRec/CDs search, round~3 asks whether the pairwise objective from round~2 helps through general emphasis on difficult examples or specifically through positive--negative separation.
It replaces the auxiliary pairwise term with focal-weighted BCE; validation NDCG@5 falls from 0.3705 to 0.3438.
The regression becomes part of the mechanism history, while the round~2 design remains the incumbent.
All four roles use this observation and the incumbent in round~4.
The selected Frontier Architect proposal restores round~2's BCE and pairwise objective and tests cosine-temperature scoring, hypothesizing that reducing embedding-norm effects will improve ranking.
The candidate implements Eq.~\ref{eq:mechanism_example} and reaches validation NDCG@5 of 0.4097.
Round~2 supplies the construction reference; round~3 supplies evidence for changing direction.

\section{Experiments}
\label{sec:experiments}

\noindent\textbf{Experimental setup.}
We study algorithm discovery from four initial recommenders---SASRec~\cite{kang2018sasrec}, MoRec~\cite{yuan2023morec}, BERT4Rec~\cite{sun2019bert4rec}, and NGCF~\cite{wang2019ngcf}---on Amazon CDs, Electronics, Office, and MovieLens-1M~\cite{harper2015movielens}.
All systems share the same preprocessed data, splits, and evaluation protocol within each model--dataset setting (Appendix~\ref{app:task-configurations}).
We compare RecRSI with the initial algorithms and five discovery systems: RecHarness, DeepEvolve, OpenEvolve, ShinkaEvolve, and Self-EvolveRec~\cite{ling2026recharness,liu2025deepevolve,sharma2025openevolve,lange2025shinkaevolve,kim2026selfevolverec}.

We reproduce all baselines from their official open-source implementations.
All systems use GPT-5.6 Terra/Luna with a common ceiling of 12 genuine evaluations, each a completed training--evaluation run, and pre-specified API-expenditure safeguards (Appendix~\ref{app:resources}).
Candidates are selected by development-set NDCG@5 and retrained for final test evaluation; results are averaged over three independent searches.
The ablation study uses MoRec on all four datasets; continuation and feedback use MoRec/CDs, and mechanism discovery uses BERT4Rec/Electronics. These analyses use development NDCG@5. Space construction compares final test performance on SASRec/CDs.
Reference API cost measures model-call expenditure, and GPU-hours measure search computation; both exclude final retraining.
Appendices~\ref{app:evaluation}--\ref{app:task-configurations} detail evaluation scope, aggregation, ranking protocols, feedback controls, and implementation settings.

\Needspace{8\baselineskip}
\subsection{Performance Comparison}
\label{sec:overall-performance}

\paragraph{Algorithm discovery across starting models.}
Table~\ref{tab:overall} reports final test performance across 16 model--dataset settings.
RecRSI improves on the initial algorithm in all 16 settings and achieves the highest mean NDCG@5 in 11 settings and the highest mean HR@5 in 12.
Averaged over all settings, the relative gains over the initial algorithms are 8.48\% in NDCG@5 and 6.88\% in HR@5; gains over the strongest competing result in each setting and metric are 2.60\% and 2.20\%.
The mean advantage remains positive after omitting any starting model or dataset (Appendix~\ref{app:head-to-head}).

BERT4Rec/Electronics links these performance gains to concrete algorithm designs.
RecRSI reaches test NDCG@5 of 0.4046, compared with 0.3881 for the initial algorithm and 0.3959 for the strongest competitor, RecHarness.
The eight selected designs combine changes to scoring, objectives, and representations.
Section~\ref{sec:discovered-mechanisms} examines how these combinations arise and which components contribute to their gains.

\begin{figure}[!htbp]
\centering
\includegraphics[width=0.96\columnwidth]{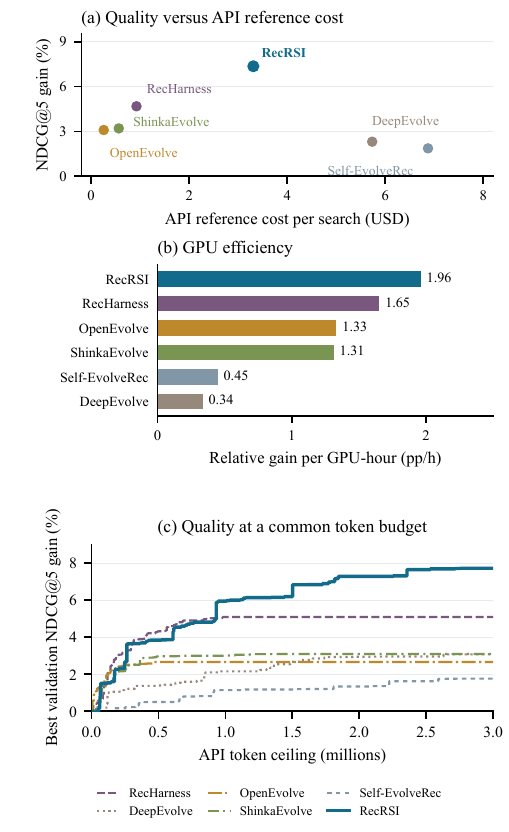}
\caption{Discovery quality and search cost. (a,b) Eight-task final test NDCG@5 gain over Seed versus API reference cost (USD) and per GPU-hour. (c) Best validation gain at a common input-plus-output token ceiling on the same eight tasks; at 3 million tokens, RecRSI reaches 7.72\% versus 5.09\% for RecHarness. Repeats are averaged within tasks, then tasks equally. Search costs exclude final retraining.}
\Description{Two panels compare final test gain with search API reference cost in USD and GPU-hours. RecRSI lies on the quality--API-cost Pareto frontier, with the highest mean test gain of 7.36 percent versus 4.68 percent for RecHarness, at 2.38 dollars more API cost per search. Its gain per GPU-hour is 1.9603. Panel c plots best validation NDCG at 5 gain against cumulative input plus output tokens from zero to three million for the same eight tasks, using the same adopted search repeats. At three million tokens, RecRSI reaches 7.7228 percent and RecHarness 5.0857 percent.}
\label{fig:discovery-cost}
\end{figure}

\paragraph{Discovery quality and research investment.}
Across the eight tasks in Figure~\ref{fig:discovery-cost}(a,b), RecRSI lies on the quality--API-cost Pareto frontier and achieves the highest average test gain.
Compared with RecHarness, it spends an additional USD~2.38 per search on average and increases the mean test NDCG@5 gain over the seed from 4.68\% to 7.36\%---a 57.17\% larger gain.
RecRSI uses additional tokens to formulate mechanism-level hypotheses, compose interventions, and integrate evidence. The higher discovery quality shows the return on this research investment.
RecRSI also leads at a common budget of 3 million tokens: mean validation gain reaches 7.72\% versus 5.09\% for RecHarness (panel c).
It achieves the highest test gain per GPU-hour, 1.96 percentage points.

\Needspace{7\baselineskip}
\subsection{Ablation Studies}
\label{sec:components}

The ablations examine how mechanism knowledge, experimental history, and research roles support algorithm construction.
Full combines specialized roles, explicit MechSpace constraints, and Mechanism Feedback.
Prompt supplies the same mechanism knowledge through natural-language guidance instead of explicit constraints; Unified uses four unspecialized roles.
Summary replaces mechanism history with a general-purpose experiment summary; Prompt+Summary applies both substitutions.
Each configuration searches MoRec on CDs, Electronics, Office, and MovieLens-1M with up to 12 evaluations. We average the relative gain of the best-so-far development score over the initial model, weighting the four datasets equally.

\begin{table}[!htbp]
\caption{Component contributions and space--feedback interaction on MoRec across four datasets.}
\label{tab:components}
\centering
\small
\setlength{\tabcolsep}{3pt}
\begin{tabular*}{\columnwidth}{@{\extracolsep{\fill}}lrr@{}}
\toprule
Method & \shortstack{Mean frontier\\gain (\%) $\uparrow$} & \shortstack{Change vs.\\Full (pp)} \\
\midrule
Full (RecRSI) & \textbf{5.6775} & -- \\
Prompt & 5.4591 & $-0.2184$ \\
Unified & 5.6387 & $-0.0388$ \\
Summary & 5.1927 & $-0.4848$ \\
Prompt + Summary & 5.5807 & $-0.0968$ \\
\bottomrule
\end{tabular*}
\par\smallskip
\begin{tabular*}{\columnwidth}{@{\extracolsep{\fill}}lrrr@{}}
\toprule
Feedback & MechSpace & Prompt & \shortstack{Space effect\\$\Delta$ (pp)} \\
\midrule
Mechanism & 5.6775 & 5.4591 & $+0.2184$ \\
Summary & 5.1927 & 5.5807 & $-0.3880$ \\
Feedback effect $\Delta$ (pp) & $+0.4848$ & $-0.1216$ & $\mathbf{+0.6064}$ \\
\bottomrule
\end{tabular*}
\par\smallskip
\begin{minipage}{\columnwidth}\footnotesize
Gains (\%) average best-development NDCG@5 improvement over the initial model equally across four datasets. Effects are percentage-point differences; bottom-right is $g_F-g_P-g_S+g_{P+S}$.
\end{minipage}
\end{table}

\paragraph{Joint value of mechanism choices and feedback.}
Full achieves the highest mean frontier gain, 5.6775\%, across the four datasets (Table~\ref{tab:components}).
With explicit MechSpace, Mechanism Feedback raises the gain from 5.1927\% under Summary to 5.6775\%, a 9.34\% increase in discovery gain.
The space--feedback interaction is positive at 0.6064 percentage points: explicit MechSpace improves the frontier with Mechanism Feedback, whereas its advantage disappears under a general summary.
The four-way comparison links the benefit of structured mechanism choices to how experimental experience guides their use.
MechSpace specifies implementable interventions; hypothesis-linked feedback connects them to the design conditions and outcomes of earlier tests, helping later hypotheses target promising combinations.

\Needspace{11\baselineskip}
\subsection{Analysis of Mechanism-Guided Research}
\label{sec:research-process}

The following studies track how earlier experiments shape later designs, from continued search to individual feedback decisions and mechanism combinations.

\subsubsection{Research Continuation vs. Restarts}
\label{sec:continuation}

Continue (Long24) extends a single search to 24 evaluations; Restart ($2\times$Short12) takes the better result from two independent 12-evaluation searches.
Both strategies receive 24 evaluation opportunities in each of three MoRec/CDs blocks.
Table~\ref{tab:continuation-endpoints} reports their final quality and API expenditure.

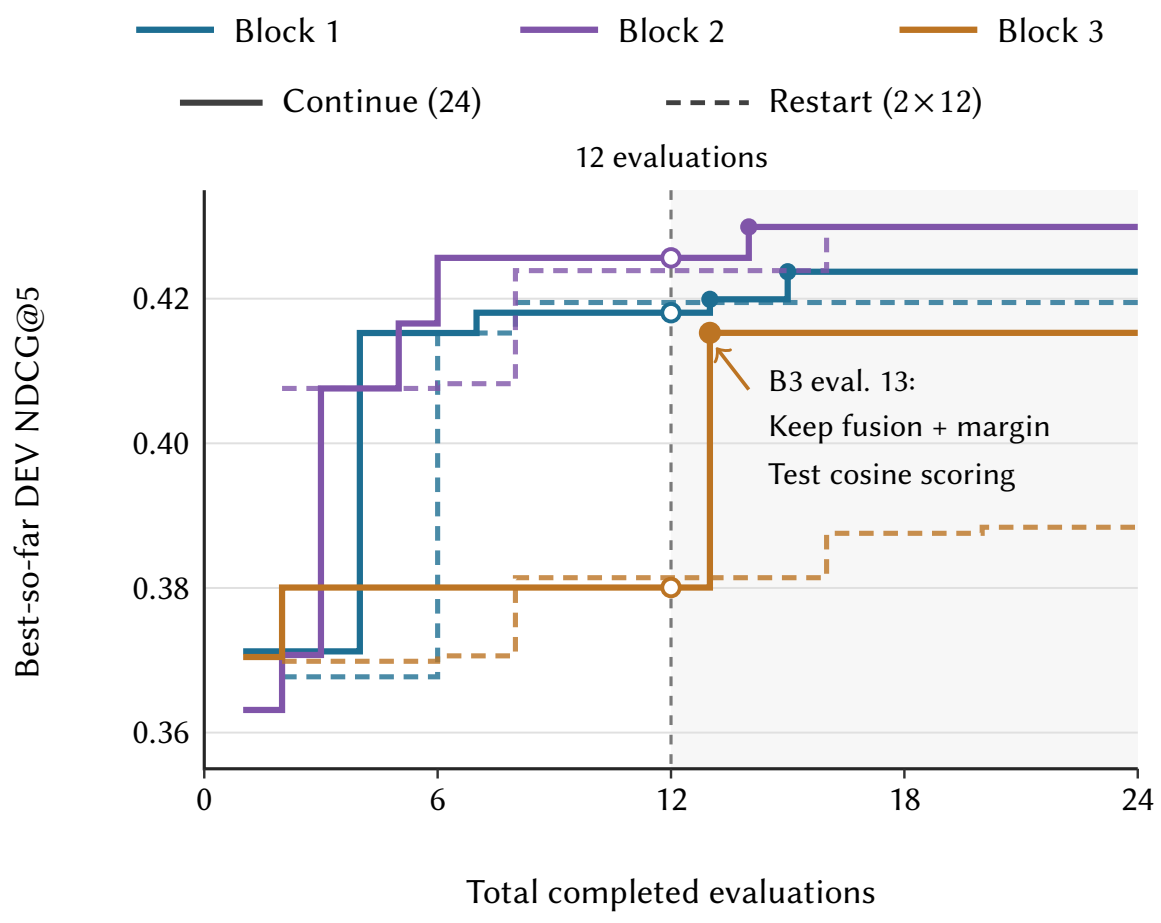
\begin{figure}[!htbp]
\centering
\resizebox{0.96\columnwidth}{!}{
\begingroup
\definecolor{rqone}{RGB}{29,110,146}
\definecolor{rqtwo}{RGB}{128,85,169}
\definecolor{rqthree}{RGB}{188,116,35}
\begin{tikzpicture}[x=1pt,y=-1pt,font=\sffamily\fontsize{7.3}{8.5}\selectfont, every node/.style={inner sep=0pt,outer sep=0pt}]
\path[use as bounding box] (0,0) rectangle (244,194);
\draw[rqone,line width=1.3pt] (29,9)--(45,9);
\node[anchor=west] at (49,9) {Block 1};
\draw[rqtwo,line width=1.3pt] (108,9)--(124,9);
\node[anchor=west] at (128,9) {Block 2};
\draw[rqthree,line width=1.3pt] (186,9)--(202,9);
\node[anchor=west] at (206,9) {Block 3};
\draw[black!75,line width=1.25pt] (38,24)--(55,24);
\node[anchor=west] at (59,24) {Continue (24)};
\draw[black!75,line width=1pt,dash pattern=on 3pt off 2pt] (138,24)--(155,24);
\node[anchor=west] at (159,24) {Restart ($2\!\times\!12$)};
\fill[black!3] (139,42) rectangle (235,161);
\draw[black!52,line width=0.65pt,dash pattern=on 2pt off 2pt] (139,42)--(139,161);
\node[anchor=center,font=\sffamily\fontsize{6.8}{7.8}\selectfont] at (139,35) {12 evaluations};
\draw[black!12,line width=0.4pt] (43.00,153.56)--(235.00,153.56);
\node[anchor=east] at (40.00,153.56) {0.36};
\draw[black!12,line width=0.4pt] (43.00,123.81)--(235.00,123.81);
\node[anchor=east] at (40.00,123.81) {0.38};
\draw[black!12,line width=0.4pt] (43.00,94.06)--(235.00,94.06);
\node[anchor=east] at (40.00,94.06) {0.40};
\draw[black!12,line width=0.4pt] (43.00,64.31)--(235.00,64.31);
\node[anchor=east] at (40.00,64.31) {0.42};
\draw[black!85,line width=0.65pt] (43,42)--(43,161)--(235,161);
\draw[black!80,line width=0.5pt] (43.000,161)--(43.000,163);
\node[anchor=north] at (43.000,165) {0};
\draw[black!80,line width=0.5pt] (91.000,161)--(91.000,163);
\node[anchor=north] at (91.000,165) {6};
\draw[black!80,line width=0.5pt] (139.000,161)--(139.000,163);
\node[anchor=north] at (139.000,165) {12};
\draw[black!80,line width=0.5pt] (187.000,161)--(187.000,163);
\node[anchor=north] at (187.000,165) {18};
\draw[black!80,line width=0.5pt] (235.000,161)--(235.000,163);
\node[anchor=north] at (235.000,165) {24};
\draw[rqone,line width=1.05pt,dash pattern=on 3pt off 2pt,opacity=0.80] (59.000,142.085)--(91.000,142.085)--(91.000,71.372)--(107.000,71.372)--(107.000,65.100)--(235.000,65.100);
\draw[rqone,line width=1.28pt] (51.000,136.852)--(75.000,136.852)--(75.000,71.372)--(99.000,71.372)--(99.000,67.198)--(147.000,67.198)--(147.000,64.443)--(163.000,64.443)--(163.000,58.771)--(235.000,58.771);
\draw[rqone,fill=white,line width=0.8pt] (139.000,67.198) circle (1.8pt);
\filldraw[rqone,line width=0.45pt] (147.000,64.443) circle (1.55pt);
\filldraw[rqone,line width=0.45pt] (163.000,58.771) circle (1.55pt);
\draw[rqtwo,line width=1.05pt,dash pattern=on 3pt off 2pt,opacity=0.80] (59.000,82.773)--(91.000,82.773)--(91.000,81.796)--(107.000,81.796)--(107.000,58.540)--(171.000,58.540)--(171.000,49.518)--(235.000,49.518);
\draw[rqtwo,line width=1.28pt] (51.000,148.903)--(59.000,148.903)--(59.000,137.591)--(67.000,137.591)--(67.000,82.773)--(83.000,82.773)--(83.000,69.411)--(91.000,69.411)--(91.000,55.913)--(155.000,55.913)--(155.000,49.554)--(235.000,49.554);
\draw[rqtwo,fill=white,line width=0.8pt] (139.000,55.913) circle (1.8pt);
\filldraw[rqtwo,line width=0.45pt] (155.000,49.554) circle (1.55pt);
\draw[rqthree,line width=1.05pt,dash pattern=on 3pt off 2pt,opacity=0.80] (59.000,138.890)--(91.000,138.890)--(91.000,137.775)--(107.000,137.775)--(107.000,121.701)--(171.000,121.701)--(171.000,112.569)--(203.000,112.569)--(203.000,111.324)--(235.000,111.324);
\draw[rqthree,line width=1.28pt] (51.000,138.013)--(59.000,138.013)--(59.000,123.722)--(147.000,123.722)--(147.000,71.338)--(235.000,71.338);
\draw[rqthree,fill=white,line width=0.8pt] (139.000,123.722) circle (1.8pt);
\filldraw[rqthree,line width=0.45pt] (147.000,71.338) circle (2.05pt);
\node[anchor=west,font=\sffamily\fontsize{6.6}{7.5}\selectfont] at (159,81) {B3 eval. 13:};
\node[anchor=west,font=\sffamily\fontsize{6.6}{7.5}\selectfont] at (159,91) {Keep fusion + margin};
\node[anchor=west,font=\sffamily\fontsize{6.6}{7.5}\selectfont] at (159,101) {Test cosine scoring};
\draw[rqthree,->,line width=0.75pt] (155,83)--(148.300,74.338);
\node[anchor=center] at (139,187) {Total completed evaluations};
\node[rotate=90,anchor=center] at (7,102) {Best-so-far DEV NDCG@5};
\end{tikzpicture}
\endgroup}
\caption{Overlaid research progress on MoRec/CDs. Colors distinguish three experiment blocks, with solid Continue and dashed Restart trajectories. At $2k$ total evaluations, Restart takes the better of the two independent searches after $k$ evaluations each. The divider marks evaluation 12; filled circles mark later bests, including the Block 3 scoring hypothesis. Table~\ref{tab:continuation-endpoints} reports final quality and API cost.}
\Description{A single plot overlays six step curves of best-so-far development NDCG at 5, sharing one total-evaluation axis. Colors denote Blocks 1--3 and solid/dashed lines denote continued research and independent restarts. After evaluation 12, Continue finds new bests at evaluations 13 and 15 in Block 1, 14 in Block 2, and 13 in Block 3. Block 3 combines the existing fusion and margin objective with cosine scoring.}
\label{fig:continuation}
\end{figure}

\begin{table}[!htbp]
\caption{Final discovery quality and continuation gains at 24 total evaluations.}
\label{tab:continuation-endpoints}
\label{tab:continuation-gains}
\centering
\small
\setlength{\tabcolsep}{3pt}
\begin{tabular*}{\columnwidth}{@{\extracolsep{\fill}}lrrrr@{}}
\toprule
Block & \shortstack{Continue\\(24)} & \shortstack{Best restart\\($2\!\times\!12$)} & \shortstack{Gain after\\eval.\ 12} & \shortstack{Continue\\$-$ restart} \\
\midrule
1 & \textbf{0.4237} & 0.4195 & $+0.0057$ & $+0.0043$ \\
2 & 0.4299 & 0.4299 & $+0.0043$ & $\approx 0$ \\
3 & \textbf{0.4153} & 0.3884 & $+0.0352$ & $+0.0269$ \\
\bottomrule
\end{tabular*}
\par\smallskip
\begin{minipage}{\columnwidth}\footnotesize
Scores and gains are development NDCG@5; differences use unrounded values (Block 2: $-2.38\!\times\!10^{-5}$).
API reference USD for Continue / both restarts: 8.81 / 8.10, 8.51 / 8.19, and 9.03 / 7.54 for Blocks 1--3.
\end{minipage}
\end{table}

Continue achieves higher final NDCG@5 in Blocks~1 and~3 and nearly ties the best restart in Block~2.
All three continued searches find new best candidates after evaluation~12 (Figure~\ref{fig:continuation}).

Block~3 shows how earlier exploration supports the later improvement.
Its best NDCG@5 remains at 0.3801 from evaluations~2 through~12.
At evaluation~13, the next hypothesis combines the existing recent-three-item fusion and margin objective with cosine-normalized scoring at temperature 0.2.
The combination reaches 0.4153, exceeding the best restart by 6.92\%.
This scoring experiment builds on the earlier representation and objective to improve the design beyond the short-search horizon.

To examine how accumulated experience guides the next design, we branch from the same MoRec/CDs incumbent with its research history available or reset.
Access to history yields a better candidate at each of three successive steps: mean NDCG@5 is 0.3947 versus 0.3681, a difference of 0.0266 (Table~\ref{tab:history-value}).
After a query-gate regression, the history-informed branch keeps the scorer and tests item-embedding gradient clipping; the reset branch switches to a pairwise objective.
Earlier evidence thus guides both the choice of a new intervention and the design it builds on (Appendix~\ref{app:history-value}).

\subsubsection{Feedback-Guided Research Decisions}
\label{sec:mechanism-feedback}

When an experiment underperforms, the next decision depends on which mechanisms to preserve and which hypothesis to test.
We compare Raw, Summary, and Mechanism at frozen decision points from real MoRec/CDs research trajectories, yielding ten next-candidate comparisons.
Raw provides the experimental records; Summary turns them into actionable research advice; Mechanism links observations to the hypotheses and mechanisms tested and proposes follow-up investigations.
Each comparison shares the research history, available actions, and downstream implementation and evaluation process; Summary and Mechanism also share the feedback model and output-token limit (Appendix~\ref{app:feedback-controls}).
Mechanism leads to higher next-candidate NDCG@5 than Summary in seven of the ten comparisons (Table~\ref{tab:feedback-decisions}).

\begin{table}[!htbp]
\centering
\small
\setlength{\tabcolsep}{4pt}
\caption{Next-candidate quality in ten feedback comparisons at frozen MoRec/CDs decision points.}
\label{tab:feedback-decisions}
\begin{tabular*}{\columnwidth}{@{\extracolsep{\fill}}lrrr@{}}
\toprule
State & Raw & Summary & Mechanism \\
\midrule
1   & 0.3647 & 0.3729 & \textbf{0.3767} \\
2   & 0.3647 & \textbf{0.3729} & \textbf{0.3729} \\
3  & 0.4123 & 0.4135 & \textbf{0.4150} \\
4   & \textbf{0.3729} & \textbf{0.3729} & \textbf{0.3729} \\
5  & 0.3681 & 0.3673 & \textbf{0.4037} \\
6   & 0.3729 & 0.3729 & \textbf{0.3755} \\
7  & 0.4079 & 0.4152 & \textbf{0.4175} \\
8   & \textbf{0.4139} & 0.3739 & 0.3996 \\
9   & 0.3339 & \textbf{0.3729} & 0.3409 \\
10  & 0.3377 & 0.3612 & \textbf{0.3625} \\
\bottomrule
\end{tabular*}
\par\smallskip
\begin{minipage}{\columnwidth}\footnotesize
Development NDCG@5; bold denotes the best value, including ties.
\end{minipage}
\end{table}

At State~6, a convolution-and-mask regression leads Mechanism Feedback to test a prefix-mean representation with weight 0.1, keeping the gate and BCE objective unchanged.
The next candidate reaches 0.3755, compared with 0.3729 under both Raw and Summary.
At State~5, Mechanism tests cosine scoring with the pairwise objective, reaching 0.4037 versus 0.3673 for Summary.
These decisions show how feedback guides both what to change and what to build on, turning the outcome of a mechanism hypothesis into the basis for the next experiment.

\subsubsection{Mechanism-Space Construction}
\label{sec:automatic-space}

We test whether MechSpace can supply effective interventions for algorithm discovery by comparing its generated space (A) with a human-designed space (H) on SASRec/CDs.

To construct H, the authors examine the initial SASRec implementation and established recommendation techniques, manually specifying mechanisms, parameter domains, and composition rules.
H defines six mechanisms, including relative-position attention, gated FFNs, and ranking objectives, with 20 single/pair structures and 95 parameter assignments.
A contains 16 fixed-parameter primitives and 494 nominal single/pair/triple combinations.
Mechanism coverage, composition rules, and space size are outcomes of the construction procedures; both resulting spaces are searched through 12 genuine evaluations.

Both searches use Research Line with the same SASRec initialization, data, evaluator, training protocol, and LLM role-to-model assignments.
Each winner is selected by validation NDCG@5, retrained with three seeds, and evaluated on held-out test data.

\begin{table}[H]
\centering
\caption{Final test performance of the validation-selected
candidates from the automatically generated (A) and
human-designed (H) mechanism spaces on SASRec/CDs.
Results are reported as mean $\pm$ sample standard deviation.}
\label{tab:space-construction}
\small
\begin{tabular*}{\columnwidth}{@{\extracolsep{\fill}}lcc@{}}
\toprule
Space & NDCG@5 & HR@5 \\
\midrule
A: MechSpace
& $\mathbf{0.4236 \pm 0.0036}$
& $\mathbf{0.5572 \pm 0.0062}$ \\
H: Human-designed
& $0.3540 \pm 0.0051$
& $0.4660 \pm 0.0053$ \\
\bottomrule
\end{tabular*}
\end{table}

\begin{table*}[!t]
\centering
\small
\setlength{\tabcolsep}{3pt}
\renewcommand{\arraystretch}{1.08}
\caption{Mechanism combinations and development-set gains in all eight BERT4Rec/Electronics searches.}
\label{tab:discovered-designs}
\begin{tabular*}{\textwidth}{@{\extracolsep{\fill}}lp{0.47\textwidth}lrrr@{}}
\toprule
Search & Selected mechanism combination & Categories & Root & Selected & Gain (\%) \\
\midrule
RO1 & Bilinear scoring + layer aggregation + label smoothing & S + A + O & 0.4079 & 0.4199 & +2.95 \\
RO2 & Cosine-temperature scoring + label smoothing & S + O & 0.4047 & \textbf{0.4398} & \textbf{+8.67} \\
RO3 & Bias-free query projection + label smoothing & S + O & 0.4045 & 0.4220 & +4.32 \\
RO4 & Shared score projection + position scaling & S + A & 0.4089 & 0.4165 & +1.87 \\
RO5 & Bounded item bias + label smoothing & S + O & 0.4054 & 0.4094 & +0.98 \\
RO6 & Relative-distance bias + masked decoder + label smoothing & R + A + O & 0.3997 & 0.4147 & +3.75 \\
RO7 & Context scoring head + mask adapter + label smoothing & S + A + O & 0.4069 & 0.4209 & +3.43 \\
RO8 & Bilinear query + deep prediction head + in-batch discrimination & S + A + O & 0.4043 & 0.4243 & +4.95 \\
\bottomrule
\end{tabular*}
\par\smallskip
\begin{minipage}{\textwidth}\footnotesize
Root and Selected are search-stage development NDCG@5; gains use each search's unrounded root value.
R: representation; A: architecture; S: scoring; O: objective. Categories describe the complete selected composition.
\end{minipage}
\end{table*}

A yields 19.65\% higher test NDCG@5 and 19.57\% higher HR@5 than H (Table~\ref{tab:space-construction}).
The selected algorithm from A normalizes embedding directions before adding positional embeddings, an intervention absent from H.
The winner from H combines bounded causal relative-position attention bias with hardness-weighted binary cross-entropy.
The comparison shows how MechSpace supports discovery: it supplies a useful intervention absent from the manual space, which Research Line turns into a higher-performing algorithm under the same evaluation budget.

\subsubsection{Discovered Mechanisms and Research Hypotheses}
\label{sec:discovered-mechanisms}

We analyze the mechanisms discovered in the three main-comparison searches (RO1--RO3) and five additional searches (RO4--RO8) on BERT4Rec/Electronics (Table~\ref{tab:discovered-designs}).
Each improves on its initial algorithm, with relative NDCG@5 gains of 0.98--8.67\% and a mean of 3.87\%.
The selected programs combine two or three mechanism categories: scoring and objective changes each appear in seven designs, architecture changes in five, and representation changes in one.
The frequent scoring--objective combinations connect the values used to learn the recommender with those used to rank items, making their compatibility a concrete question for mechanism-guided research.

RO2 produces the largest gain, 8.67\%, by combining cosine-temperature scoring with label-smoothed training while keeping the encoder unchanged.
The shared scorer coordinates training logits and item ranking.
Component comparisons for RO1--RO3 (Appendix~\ref{app:bert-case}) expose why design context matters.
With bilinear scoring fixed in RO1, layer aggregation changes NDCG@5 by $-0.0014$ without smoothing but $+0.0015$ with it.
This component comparison identifies an objective-dependent design opportunity: layer aggregation improves the design when paired with smoothing.
A hypothesis records the intervention together with its surrounding design so that this dependence can guide the next experiment.
In RO2, smoothing adds 0.0019 to an already effective cosine scorer.

RO3's pre-experiment hypothesis shows how prior results shape the next combination.
It builds on label smoothing, which had improved development NDCG@5 from 0.4045 to 0.4101, and uses the regression of a residual-head alternative to motivate a new scoring hypothesis.
The improved, smoothed design becomes the reference for testing the new scorer.
A bias-free query projection tests whether learned matching geometry can complement this objective; the resulting design reaches 0.4220.
The combination exceeds both projection alone (0.4170) and smoothing alone (0.4101).
The proposal history explains why the scorer was tested; the component comparisons show what it adds to the design.

\section{Conclusion}
\label{sec:conclusion}

RecRSI makes mechanism hypotheses the unit of cumulative recommendation research.
Its task-derived mechanism space, reference-based candidate construction, and mechanism feedback allow research to build on promising designs and learn from both successful and unsuccessful experiments.
Across 16 model--dataset settings, RecRSI improves both ranking metrics over every initial recommender and achieves positive average gains over the strongest competing results.
The process analyses show how accumulated evidence redirects research and develops stronger mechanism combinations.
The resulting research cycle uses each experiment to inform the next mechanism choice.

\FloatBarrier
\bibliographystyle{ACM-Reference-Format}
\bibliography{references,references_added}


\begin{thebibliography}{38}


\ifx \showCODEN    \undefined \def \showCODEN     #1{\unskip}     \fi
\ifx \showISBNx    \undefined \def \showISBNx     #1{\unskip}     \fi
\ifx \showISBNxiii \undefined \def \showISBNxiii  #1{\unskip}     \fi
\ifx \showISSN     \undefined \def \showISSN      #1{\unskip}     \fi
\ifx \showLCCN     \undefined \def \showLCCN      #1{\unskip}     \fi
\ifx \shownote     \undefined \def \shownote      #1{#1}          \fi
\ifx \showarticletitle \undefined \def \showarticletitle #1{#1}   \fi
\ifx \showURL      \undefined \def \showURL       {\relax}        \fi
\providecommand\bibfield[2]{#2}
\providecommand\bibinfo[2]{#2}
\providecommand\natexlab[1]{#1}
\providecommand\showeprint[2][]{arXiv:#2}

\bibitem[Chen et~al\mbox{.}(2026)]%
        {chen2026causalevolve}
\bibfield{author}{\bibinfo{person}{Yongqiang Chen}, \bibinfo{person}{Chenxi
  Liu}, \bibinfo{person}{Zhenhao Chen}, \bibinfo{person}{Tongliang Liu},
  \bibinfo{person}{Bo Han}, {and} \bibinfo{person}{Kun Zhang}.}
  \bibinfo{year}{2026}\natexlab{}.
\newblock \bibinfo{title}{{CausalEvolve}: Towards Open-Ended Discovery with
  Causal Scratchpad}.
\newblock
\showeprint[arxiv]{2603.14575}
\href{https://doi.org/10.48550/arXiv.2603.14575}{doi:\nolinkurl{10.48550/arXiv.2603.14575}}


\bibitem[Ferrari~Dacrema et~al\mbox{.}(2019)]%
        {dacrema2019progress}
\bibfield{author}{\bibinfo{person}{Maurizio Ferrari~Dacrema},
  \bibinfo{person}{Paolo Cremonesi}, {and} \bibinfo{person}{Dietmar Jannach}.}
  \bibinfo{year}{2019}\natexlab{}.
\newblock \showarticletitle{Are We Really Making Much Progress? A Worrying
  Analysis of Recent Neural Recommendation Approaches}. In
  \bibinfo{booktitle}{\emph{Proceedings of RecSys}}. \bibinfo{publisher}{ACM},
  \bibinfo{address}{New York, NY, USA}, \bibinfo{pages}{101--109}.
\newblock
\href{https://doi.org/10.1145/3298689.3347058}{doi:\nolinkurl{10.1145/3298689.3347058}}


\bibitem[Harper and Konstan(2015)]%
        {harper2015movielens}
\bibfield{author}{\bibinfo{person}{F.~Maxwell Harper} {and}
  \bibinfo{person}{Joseph~A. Konstan}.} \bibinfo{year}{2015}\natexlab{}.
\newblock \showarticletitle{The {MovieLens} Datasets: History and Context}.
\newblock \bibinfo{journal}{\emph{ACM Transactions on Interactive Intelligent
  Systems}} \bibinfo{volume}{5}, \bibinfo{number}{4}, Article
  \bibinfo{articleno}{19} (\bibinfo{year}{2015}), \bibinfo{numpages}{19}~pages.
\newblock
\href{https://doi.org/10.1145/2827872}{doi:\nolinkurl{10.1145/2827872}}


\bibitem[He et~al\mbox{.}(2020)]%
        {he2020lightgcn}
\bibfield{author}{\bibinfo{person}{Xiangnan He}, \bibinfo{person}{Kuan Deng},
  \bibinfo{person}{Xiang Wang}, \bibinfo{person}{Yan Li},
  \bibinfo{person}{Yongdong Zhang}, {and} \bibinfo{person}{Meng Wang}.}
  \bibinfo{year}{2020}\natexlab{}.
\newblock \showarticletitle{{LightGCN}: Simplifying and Powering Graph
  Convolution Network for Recommendation}. In
  \bibinfo{booktitle}{\emph{Proceedings of SIGIR}}. \bibinfo{publisher}{ACM},
  \bibinfo{address}{New York, NY, USA}, \bibinfo{pages}{639--648}.
\newblock
\href{https://doi.org/10.1145/3397271.3401063}{doi:\nolinkurl{10.1145/3397271.3401063}}


\bibitem[Jeong et~al\mbox{.}(2026)]%
        {jeong2026agentnas}
\bibfield{author}{\bibinfo{person}{Seokhoon Jeong}, \bibinfo{person}{Mijung
  Kim}, {and} \bibinfo{person}{Taehwan Kim}.} \bibinfo{year}{2026}\natexlab{}.
\newblock \bibinfo{title}{Agentic Neural Architecture Search}.
\newblock
\showeprint[arxiv]{2607.07984}
\href{https://doi.org/10.48550/arXiv.2607.07984}{doi:\nolinkurl{10.48550/arXiv.2607.07984}}


\bibitem[Jiang et~al\mbox{.}(2025)]%
        {jiang2025aide}
\bibfield{author}{\bibinfo{person}{Zhengyao Jiang}, \bibinfo{person}{Dominik
  Schmidt}, \bibinfo{person}{Dhruv Srikanth}, \bibinfo{person}{Dixing Xu},
  \bibinfo{person}{Ian Kaplan}, \bibinfo{person}{Deniss Jacenko}, {and}
  \bibinfo{person}{Yuxiang Wu}.} \bibinfo{year}{2025}\natexlab{}.
\newblock \bibinfo{title}{{AIDE}: {AI}-Driven Exploration in the Space of
  Code}.
\newblock
\showeprint[arxiv]{2502.13138}
\href{https://doi.org/10.48550/arXiv.2502.13138}{doi:\nolinkurl{10.48550/arXiv.2502.13138}}


\bibitem[Kang and McAuley(2018)]%
        {kang2018sasrec}
\bibfield{author}{\bibinfo{person}{Wang-Cheng Kang} {and}
  \bibinfo{person}{Julian McAuley}.} \bibinfo{year}{2018}\natexlab{}.
\newblock \showarticletitle{Self-Attentive Sequential Recommendation}. In
  \bibinfo{booktitle}{\emph{2018 IEEE International Conference on Data Mining
  (ICDM)}}. \bibinfo{publisher}{IEEE}, \bibinfo{address}{Piscataway, NJ, USA},
  \bibinfo{pages}{197--206}.
\newblock
\href{https://doi.org/10.1109/ICDM.2018.00035}{doi:\nolinkurl{10.1109/ICDM.2018.00035}}


\bibitem[Karpathy(2026)]%
        {karpathy2026autoresearch}
\bibfield{author}{\bibinfo{person}{Andrej Karpathy}.}
  \bibinfo{year}{2026}\natexlab{}.
\newblock \bibinfo{booktitle}{\emph{{autoresearch}: {AI} Agents Running
  Research on Single-{GPU} Nanochat Training Automatically}}.
\newblock GitHub.
\newblock
\urldef\tempurl%
\url{https://github.com/karpathy/autoresearch}
\showURL{%
\tempurl}


\bibitem[Kim et~al\mbox{.}(2026)]%
        {kim2026selfevolverec}
\bibfield{author}{\bibinfo{person}{Sein Kim}, \bibinfo{person}{Sangwu Park},
  \bibinfo{person}{Hongseok Kang}, \bibinfo{person}{Wonjoong Kim},
  \bibinfo{person}{Jimin Seo}, \bibinfo{person}{Yeonjun In},
  \bibinfo{person}{Kanghoon Yoon}, \bibinfo{person}{Hyunsik Jeon}, {and}
  \bibinfo{person}{Chanyoung Park}.} \bibinfo{year}{2026}\natexlab{}.
\newblock \bibinfo{title}{{Self-EvolveRec}: Self-Evolving Recommender Systems
  with {LLM}-Based Directional Feedback}.
\newblock
\showeprint[arxiv]{2602.12612}
\href{https://doi.org/10.48550/arXiv.2602.12612}{doi:\nolinkurl{10.48550/arXiv.2602.12612}}


\bibitem[Lange et~al\mbox{.}(2026)]%
        {lange2025shinkaevolve}
\bibfield{author}{\bibinfo{person}{Robert~Tjarko Lange}, \bibinfo{person}{Yuki
  Imajuku}, {and} \bibinfo{person}{Edoardo Cetin}.}
  \bibinfo{year}{2026}\natexlab{}.
\newblock \showarticletitle{{ShinkaEvolve}: Towards Open-Ended and
  Sample-Efficient Program Evolution}. In \bibinfo{booktitle}{\emph{The
  Fourteenth International Conference on Learning Representations}}.
  \bibinfo{pages}{74026--74078}.
\newblock
\urldef\tempurl%
\url{https://proceedings.iclr.cc/paper_files/paper/2026/hash/7886b9bafe76c52fd568db10ff9772df-Abstract-Conference.html}
\showURL{%
\tempurl}


\bibitem[Lin et~al\mbox{.}(2025)]%
        {lin2025llmrecsurvey}
\bibfield{author}{\bibinfo{person}{Jianghao Lin}, \bibinfo{person}{Xinyi Dai},
  \bibinfo{person}{Yunjia Xi}, \bibinfo{person}{Weiwen Liu},
  \bibinfo{person}{Bo Chen}, \bibinfo{person}{Hao Zhang}, \bibinfo{person}{Yong
  Liu}, \bibinfo{person}{Chuhan Wu}, \bibinfo{person}{Xiangyang Li},
  \bibinfo{person}{Chenxu Zhu}, \bibinfo{person}{Huifeng Guo},
  \bibinfo{person}{Yong Yu}, \bibinfo{person}{Ruiming Tang}, {and}
  \bibinfo{person}{Weinan Zhang}.} \bibinfo{year}{2025}\natexlab{}.
\newblock \showarticletitle{How Can Recommender Systems Benefit from Large
  Language Models: A Survey}.
\newblock \bibinfo{journal}{\emph{ACM Transactions on Information Systems}}
  \bibinfo{volume}{43}, \bibinfo{number}{2}, Article \bibinfo{articleno}{28}
  (\bibinfo{year}{2025}), \bibinfo{numpages}{47}~pages.
\newblock
\href{https://doi.org/10.1145/3678004}{doi:\nolinkurl{10.1145/3678004}}


\bibitem[Ling et~al\mbox{.}(2026)]%
        {ling2026recharness}
\bibfield{author}{\bibinfo{person}{Haoran Ling}, \bibinfo{person}{Yuecheng Li},
  \bibinfo{person}{Zeyu Song}, \bibinfo{person}{Jing Yao},
  \bibinfo{person}{Shuwen Kang}, \bibinfo{person}{Chi Lu},
  \bibinfo{person}{Wenjin Wu}, {and} \bibinfo{person}{Peng Jiang}.}
  \bibinfo{year}{2026}\natexlab{}.
\newblock \bibinfo{title}{{RecHarness}: A Bandit-Routed Agentic Harness for
  Self-Evolving Recommender Systems}.
\newblock
\showeprint[arxiv]{2607.29241}
\urldef\tempurl%
\url{https://arxiv.org/abs/2607.29241}
\showURL{%
\tempurl}


\bibitem[Liu et~al\mbox{.}(2020)]%
        {liu2020autofis}
\bibfield{author}{\bibinfo{person}{Bin Liu}, \bibinfo{person}{Chenxu Zhu},
  \bibinfo{person}{Guilin Li}, \bibinfo{person}{Weinan Zhang},
  \bibinfo{person}{Jincai Lai}, \bibinfo{person}{Ruiming Tang},
  \bibinfo{person}{Xiuqiang He}, \bibinfo{person}{Zhenguo Li}, {and}
  \bibinfo{person}{Yong Yu}.} \bibinfo{year}{2020}\natexlab{}.
\newblock \showarticletitle{{AutoFIS}: Automatic Feature Interaction Selection
  in Factorization Models for Click-Through Rate Prediction}. In
  \bibinfo{booktitle}{\emph{Proceedings of KDD}}. \bibinfo{publisher}{ACM},
  \bibinfo{address}{New York, NY, USA}, \bibinfo{pages}{2636--2645}.
\newblock
\href{https://doi.org/10.1145/3394486.3403314}{doi:\nolinkurl{10.1145/3394486.3403314}}


\bibitem[Liu et~al\mbox{.}(2024)]%
        {liu2024eoh}
\bibfield{author}{\bibinfo{person}{Fei Liu}, \bibinfo{person}{Xialiang Tong},
  \bibinfo{person}{Mingxuan Yuan}, \bibinfo{person}{Xi Lin},
  \bibinfo{person}{Fu Luo}, \bibinfo{person}{Zhenkun Wang},
  \bibinfo{person}{Zhichao Lu}, {and} \bibinfo{person}{Qingfu Zhang}.}
  \bibinfo{year}{2024}\natexlab{}.
\newblock \showarticletitle{Evolution of Heuristics: Towards Efficient
  Automatic Algorithm Design Using Large Language Model}. In
  \bibinfo{booktitle}{\emph{Proceedings of the 41st International Conference on
  Machine Learning}} \emph{(\bibinfo{series}{Proceedings of Machine Learning
  Research}, Vol.~\bibinfo{volume}{235})}. \bibinfo{publisher}{PMLR},
  \bibinfo{address}{Cambridge, MA, USA}, \bibinfo{pages}{32201--32223}.
\newblock
\urldef\tempurl%
\url{https://proceedings.mlr.press/v235/liu24bs.html}
\showURL{%
\tempurl}


\bibitem[Liu et~al\mbox{.}(2025)]%
        {liu2025deepevolve}
\bibfield{author}{\bibinfo{person}{Gang Liu}, \bibinfo{person}{Yihan Zhu},
  \bibinfo{person}{Jie Chen}, {and} \bibinfo{person}{Meng Jiang}.}
  \bibinfo{year}{2025}\natexlab{}.
\newblock \bibinfo{title}{Scientific Algorithm Discovery by Augmenting
  {AlphaEvolve} with Deep Research}.
\newblock
\showeprint[arxiv]{2510.06056}
\href{https://doi.org/10.48550/arXiv.2510.06056}{doi:\nolinkurl{10.48550/arXiv.2510.06056}}


\bibitem[Lu et~al\mbox{.}(2024)]%
        {lu2024aiscientist}
\bibfield{author}{\bibinfo{person}{Chris Lu}, \bibinfo{person}{Cong Lu},
  \bibinfo{person}{Robert~Tjarko Lange}, \bibinfo{person}{Jakob Foerster},
  \bibinfo{person}{Jeff Clune}, {and} \bibinfo{person}{David Ha}.}
  \bibinfo{year}{2024}\natexlab{}.
\newblock \bibinfo{title}{The {AI Scientist}: Towards Fully Automated
  Open-Ended Scientific Discovery}.
\newblock
\showeprint[arxiv]{2408.06292}
\href{https://doi.org/10.48550/arXiv.2408.06292}{doi:\nolinkurl{10.48550/arXiv.2408.06292}}


\bibitem[Madaan et~al\mbox{.}(2023)]%
        {madaan2023selfrefine}
\bibfield{author}{\bibinfo{person}{Aman Madaan}, \bibinfo{person}{Niket
  Tandon}, \bibinfo{person}{Prakhar Gupta}, \bibinfo{person}{Skyler Hallinan},
  \bibinfo{person}{Luyu Gao}, \bibinfo{person}{Sarah Wiegreffe},
  \bibinfo{person}{Uri Alon}, \bibinfo{person}{Nouha Dziri},
  \bibinfo{person}{Shrimai Prabhumoye}, \bibinfo{person}{Yiming Yang},
  \bibinfo{person}{Shashank Gupta}, \bibinfo{person}{Bodhisattwa~Prasad
  Majumder}, \bibinfo{person}{Katherine Hermann}, \bibinfo{person}{Sean
  Welleck}, \bibinfo{person}{Amir Yazdanbakhsh}, {and} \bibinfo{person}{Peter
  Clark}.} \bibinfo{year}{2023}\natexlab{}.
\newblock \showarticletitle{{Self-Refine}: Iterative Refinement with
  Self-Feedback}. In \bibinfo{booktitle}{\emph{Advances in Neural Information
  Processing Systems}}, Vol.~\bibinfo{volume}{36}. \bibinfo{publisher}{Curran
  Associates, Inc.}, \bibinfo{address}{Red Hook, NY, USA},
  \bibinfo{pages}{46534--46594}.
\newblock
\href{https://doi.org/10.52202/075280-2019}{doi:\nolinkurl{10.52202/075280-2019}}


\bibitem[Mao et~al\mbox{.}(2021)]%
        {mao2021simplex}
\bibfield{author}{\bibinfo{person}{Kelong Mao}, \bibinfo{person}{Jieming Zhu},
  \bibinfo{person}{Jinpeng Wang}, \bibinfo{person}{Quanyu Dai},
  \bibinfo{person}{Zhenhua Dong}, \bibinfo{person}{Xi Xiao}, {and}
  \bibinfo{person}{Xiuqiang He}.} \bibinfo{year}{2021}\natexlab{}.
\newblock \showarticletitle{{SimpleX}: A Simple and Strong Baseline for
  Collaborative Filtering}. In \bibinfo{booktitle}{\emph{Proceedings of the
  30th ACM International Conference on Information and Knowledge Management}}
  \emph{(\bibinfo{series}{CIKM '21})}. \bibinfo{publisher}{Association for
  Computing Machinery}, \bibinfo{address}{New York, NY, USA},
  \bibinfo{pages}{1243--1252}.
\newblock
\href{https://doi.org/10.1145/3459637.3482297}{doi:\nolinkurl{10.1145/3459637.3482297}}


\bibitem[McAuley et~al\mbox{.}(2015)]%
        {mcauley2015styles}
\bibfield{author}{\bibinfo{person}{Julian McAuley},
  \bibinfo{person}{Christopher Targett}, \bibinfo{person}{Qinfeng Shi}, {and}
  \bibinfo{person}{Anton van~den Hengel}.} \bibinfo{year}{2015}\natexlab{}.
\newblock \showarticletitle{Image-based Recommendations on Styles and
  Substitutes}. In \bibinfo{booktitle}{\emph{Proceedings of the 38th
  International ACM SIGIR Conference on Research and Development in Information
  Retrieval}}. \bibinfo{publisher}{Association for Computing Machinery},
  \bibinfo{address}{New York, NY, USA}, \bibinfo{pages}{43--52}.
\newblock
\href{https://doi.org/10.1145/2766462.2767755}{doi:\nolinkurl{10.1145/2766462.2767755}}


\bibitem[Mu et~al\mbox{.}(2026)]%
        {mu2026evorec}
\bibfield{author}{\bibinfo{person}{Lingyu Mu}, \bibinfo{person}{Hao Deng},
  \bibinfo{person}{Haibo Xing}, \bibinfo{person}{Jinxin Hu},
  \bibinfo{person}{Yu Zhang}, {and} \bibinfo{person}{Xiaoyi Zeng}.}
  \bibinfo{year}{2026}\natexlab{}.
\newblock \bibinfo{title}{{EvoRec}: Self Evolving Agentic Recommender Systems}.
\newblock
\showeprint[arxiv]{2606.28368}
\href{https://doi.org/10.48550/arXiv.2606.28368}{doi:\nolinkurl{10.48550/arXiv.2606.28368}}


\bibitem[Novikov et~al\mbox{.}(2025)]%
        {novikov2025alphaevolve}
\bibfield{author}{\bibinfo{person}{Alexander Novikov},
  \bibinfo{person}{Ng{\^a}n V{\~u}}, \bibinfo{person}{Marvin Eisenberger},
  \bibinfo{person}{Emilien Dupont}, \bibinfo{person}{Po-Sen Huang},
  \bibinfo{person}{Adam~Zsolt Wagner}, \bibinfo{person}{Sergey Shirobokov},
  \bibinfo{person}{Borislav Kozlovskii}, \bibinfo{person}{Francisco J.~R.
  Ruiz}, \bibinfo{person}{Abbas Mehrabian}, \bibinfo{person}{M.~Pawan Kumar},
  \bibinfo{person}{Abigail See}, \bibinfo{person}{Swarat Chaudhuri},
  \bibinfo{person}{George Holland}, \bibinfo{person}{Alex Davies},
  \bibinfo{person}{Sebastian Nowozin}, \bibinfo{person}{Pushmeet Kohli}, {and}
  \bibinfo{person}{Matej Balog}.} \bibinfo{year}{2025}\natexlab{}.
\newblock \bibinfo{title}{{AlphaEvolve}: A Coding Agent for Scientific and
  Algorithmic Discovery}.
\newblock
\showeprint[arxiv]{2506.13131}
\href{https://doi.org/10.48550/arXiv.2506.13131}{doi:\nolinkurl{10.48550/arXiv.2506.13131}}


\bibitem[Real et~al\mbox{.}(2020)]%
        {real2020automlzero}
\bibfield{author}{\bibinfo{person}{Esteban Real}, \bibinfo{person}{Chen Liang},
  \bibinfo{person}{David So}, {and} \bibinfo{person}{Quoc Le}.}
  \bibinfo{year}{2020}\natexlab{}.
\newblock \showarticletitle{{AutoML-Zero}: Evolving Machine Learning Algorithms
  From Scratch}. In \bibinfo{booktitle}{\emph{Proceedings of ICML}}
  \emph{(\bibinfo{series}{Proceedings of Machine Learning Research},
  Vol.~\bibinfo{volume}{119})}. \bibinfo{publisher}{PMLR},
  \bibinfo{address}{Cambridge, MA, USA}, \bibinfo{pages}{8007--8019}.
\newblock
\urldef\tempurl%
\url{https://proceedings.mlr.press/v119/real20a.html}
\showURL{%
\tempurl}


\bibitem[Rendle et~al\mbox{.}(2020)]%
        {rendle2020revisited}
\bibfield{author}{\bibinfo{person}{Steffen Rendle}, \bibinfo{person}{Walid
  Krichene}, \bibinfo{person}{Li Zhang}, {and} \bibinfo{person}{John~Roberts
  Anderson}.} \bibinfo{year}{2020}\natexlab{}.
\newblock \showarticletitle{Neural Collaborative Filtering vs. Matrix
  Factorization Revisited}. In \bibinfo{booktitle}{\emph{Proceedings of the
  14th ACM Conference on Recommender Systems}}. \bibinfo{publisher}{Association
  for Computing Machinery}, \bibinfo{address}{New York, NY, USA},
  \bibinfo{pages}{240--248}.
\newblock
\href{https://doi.org/10.1145/3383313.3412488}{doi:\nolinkurl{10.1145/3383313.3412488}}


\bibitem[Romera-Paredes et~al\mbox{.}(2024)]%
        {romeraparedes2024funsearch}
\bibfield{author}{\bibinfo{person}{Bernardino Romera-Paredes},
  \bibinfo{person}{Mohammadamin Barekatain}, \bibinfo{person}{Alexander
  Novikov}, \bibinfo{person}{Matej Balog}, \bibinfo{person}{M.~Pawan Kumar},
  \bibinfo{person}{Emilien Dupont}, \bibinfo{person}{Francisco J.~R. Ruiz},
  \bibinfo{person}{Jordan~S. Ellenberg}, \bibinfo{person}{Pengming Wang},
  \bibinfo{person}{Omar Fawzi}, \bibinfo{person}{Pushmeet Kohli}, {and}
  \bibinfo{person}{Alhussein Fawzi}.} \bibinfo{year}{2024}\natexlab{}.
\newblock \showarticletitle{Mathematical Discoveries from Program Search with
  Large Language Models}.
\newblock \bibinfo{journal}{\emph{Nature}}  \bibinfo{volume}{625}
  (\bibinfo{year}{2024}), \bibinfo{pages}{468--475}.
\newblock
\href{https://doi.org/10.1038/s41586-023-06924-6}{doi:\nolinkurl{10.1038/s41586-023-06924-6}}


\bibitem[Sakuma et~al\mbox{.}(2026)]%
        {sakuma2026fairnad}
\bibfield{author}{\bibinfo{person}{Yuiko Sakuma}, \bibinfo{person}{Masakazu
  Yoshimura}, \bibinfo{person}{Marcel Gr{\"o}pl}, \bibinfo{person}{Zitang Sun},
  \bibinfo{person}{Junji Otsuka}, \bibinfo{person}{Atsushi Irie}, {and}
  \bibinfo{person}{Takeshi Ohashi}.} \bibinfo{year}{2026}\natexlab{}.
\newblock \bibinfo{title}{Structuring Open-Ended {NAS}: Semi-Automated Design
  Knowledge Structuring with {LLMs} for Efficient Neural Architecture Search}.
\newblock
\showeprint[arxiv]{2605.19247}
\href{https://doi.org/10.48550/arXiv.2605.19247}{doi:\nolinkurl{10.48550/arXiv.2605.19247}}


\bibitem[Sharma(2025)]%
        {sharma2025openevolve}
\bibfield{author}{\bibinfo{person}{Asankhaya Sharma}.}
  \bibinfo{year}{2025}\natexlab{}.
\newblock \bibinfo{booktitle}{\emph{{OpenEvolve}: An Open-Source Evolutionary
  Coding Agent}}.
\newblock GitHub.
\newblock
\urldef\tempurl%
\url{https://github.com/algorithmicsuperintelligence/openevolve}
\showURL{%
\tempurl}


\bibitem[Shinn et~al\mbox{.}(2023)]%
        {shinn2023reflexion}
\bibfield{author}{\bibinfo{person}{Noah Shinn}, \bibinfo{person}{Federico
  Cassano}, \bibinfo{person}{Ashwin Gopinath}, \bibinfo{person}{Karthik
  Narasimhan}, {and} \bibinfo{person}{Shunyu Yao}.}
  \bibinfo{year}{2023}\natexlab{}.
\newblock \showarticletitle{{Reflexion}: Language Agents with Verbal
  Reinforcement Learning}. In \bibinfo{booktitle}{\emph{Advances in Neural
  Information Processing Systems}}, Vol.~\bibinfo{volume}{36}.
  \bibinfo{publisher}{Curran Associates, Inc.}, \bibinfo{address}{Red Hook, NY,
  USA}, \bibinfo{pages}{8634--8652}.
\newblock
\urldef\tempurl%
\url{https://proceedings.neurips.cc/paper_files/paper/2023/hash/1b44b878bb782e6954cd888628510e90-Abstract-Conference.html}
\showURL{%
\tempurl}


\bibitem[Song et~al\mbox{.}(2020)]%
        {song2020autoctr}
\bibfield{author}{\bibinfo{person}{Qingquan Song}, \bibinfo{person}{Dehua
  Cheng}, \bibinfo{person}{Hanning Zhou}, \bibinfo{person}{Jiyan Yang},
  \bibinfo{person}{Yuandong Tian}, {and} \bibinfo{person}{Xia Hu}.}
  \bibinfo{year}{2020}\natexlab{}.
\newblock \showarticletitle{Towards Automated Neural Interaction Discovery for
  Click-Through Rate Prediction}. In \bibinfo{booktitle}{\emph{Proceedings of
  KDD}}. \bibinfo{publisher}{ACM}, \bibinfo{address}{New York, NY, USA},
  \bibinfo{pages}{945--955}.
\newblock
\href{https://doi.org/10.1145/3394486.3403137}{doi:\nolinkurl{10.1145/3394486.3403137}}


\bibitem[Sun et~al\mbox{.}(2019)]%
        {sun2019bert4rec}
\bibfield{author}{\bibinfo{person}{Fei Sun}, \bibinfo{person}{Jun Liu},
  \bibinfo{person}{Jian Wu}, \bibinfo{person}{Changhua Pei},
  \bibinfo{person}{Xiao Lin}, \bibinfo{person}{Wenwu Ou}, {and}
  \bibinfo{person}{Peng Jiang}.} \bibinfo{year}{2019}\natexlab{}.
\newblock \showarticletitle{{BERT4Rec}: Sequential Recommendation with
  Bidirectional Encoder Representations from Transformer}. In
  \bibinfo{booktitle}{\emph{Proceedings of the 28th ACM International
  Conference on Information and Knowledge Management}}.
  \bibinfo{publisher}{Association for Computing Machinery},
  \bibinfo{address}{New York, NY, USA}, \bibinfo{pages}{1441--1450}.
\newblock
\href{https://doi.org/10.1145/3357384.3357895}{doi:\nolinkurl{10.1145/3357384.3357895}}


\bibitem[Wang et~al\mbox{.}(2019)]%
        {wang2019ngcf}
\bibfield{author}{\bibinfo{person}{Xiang Wang}, \bibinfo{person}{Xiangnan He},
  \bibinfo{person}{Meng Wang}, \bibinfo{person}{Fuli Feng}, {and}
  \bibinfo{person}{Tat-Seng Chua}.} \bibinfo{year}{2019}\natexlab{}.
\newblock \showarticletitle{Neural Graph Collaborative Filtering}. In
  \bibinfo{booktitle}{\emph{Proceedings of the 42nd International ACM SIGIR
  Conference on Research and Development in Information Retrieval}}.
  \bibinfo{publisher}{Association for Computing Machinery},
  \bibinfo{address}{New York, NY, USA}, \bibinfo{pages}{165--174}.
\newblock
\href{https://doi.org/10.1145/3331184.3331267}{doi:\nolinkurl{10.1145/3331184.3331267}}


\bibitem[Yamada et~al\mbox{.}(2025)]%
        {yamada2025aiscientistv2}
\bibfield{author}{\bibinfo{person}{Yutaro Yamada},
  \bibinfo{person}{Robert~Tjarko Lange}, \bibinfo{person}{Cong Lu},
  \bibinfo{person}{Shengran Hu}, \bibinfo{person}{Chris Lu},
  \bibinfo{person}{Jakob Foerster}, \bibinfo{person}{Jeff Clune}, {and}
  \bibinfo{person}{David Ha}.} \bibinfo{year}{2025}\natexlab{}.
\newblock \bibinfo{title}{The {AI Scientist-v2}: Workshop-Level Automated
  Scientific Discovery via Agentic Tree Search}.
\newblock
\showeprint[arxiv]{2504.08066}
\href{https://doi.org/10.48550/arXiv.2504.08066}{doi:\nolinkurl{10.48550/arXiv.2504.08066}}


\bibitem[Yang et~al\mbox{.}(2026)]%
        {yang2026agentxmodel}
\bibfield{author}{\bibinfo{person}{Shuang Yang}, \bibinfo{person}{Zijie
  Zhuang}, \bibinfo{person}{Changxin Lao}, \bibinfo{person}{Pengbo Xu},
  \bibinfo{person}{Hanwen Xu}, \bibinfo{person}{Yusheng Huang},
  \bibinfo{person}{Han Gao}, \bibinfo{person}{Guanchen Wang},
  \bibinfo{person}{Tianbao Ma}, \bibinfo{person}{Linxun Chen},
  \bibinfo{person}{Peilin Song}, \bibinfo{person}{Xuming Wang},
  \bibinfo{person}{Chen Li}, \bibinfo{person}{Fan Wu}, \bibinfo{person}{Tao
  Wang}, \bibinfo{person}{Zibo Zhao}, \bibinfo{person}{Xiangyu Wu},
  \bibinfo{person}{An Liu}, \bibinfo{person}{Fei Pan}, \bibinfo{person}{Peng
  Jiang}, \bibinfo{person}{Chen Yang}, \bibinfo{person}{Zhaojie Liu}, {and}
  \bibinfo{person}{Wenwu Ou}.} \bibinfo{year}{2026}\natexlab{}.
\newblock \bibinfo{title}{Advancing Model Research in {AgentX}: Long-Horizon
  Autonomy for Industrial Recommender Systems}.
\newblock
\showeprint[arxiv]{2609.30001}
\urldef\tempurl%
\url{https://arxiv.org/abs/2609.30001}
\showURL{%
\tempurl}


\bibitem[Ye et~al\mbox{.}(2024)]%
        {ye2024reevo}
\bibfield{author}{\bibinfo{person}{Haoran Ye}, \bibinfo{person}{Jiarui Wang},
  \bibinfo{person}{Zhiguang Cao}, \bibinfo{person}{Federico Berto},
  \bibinfo{person}{Chuanbo Hua}, \bibinfo{person}{Haeyeon Kim},
  \bibinfo{person}{Jinkyoo Park}, {and} \bibinfo{person}{Guojie Song}.}
  \bibinfo{year}{2024}\natexlab{}.
\newblock \showarticletitle{{ReEvo}: Large Language Models as Hyper-Heuristics
  with Reflective Evolution}. In \bibinfo{booktitle}{\emph{Advances in Neural
  Information Processing Systems}}, Vol.~\bibinfo{volume}{37}.
  \bibinfo{publisher}{Curran Associates, Inc.}, \bibinfo{address}{Red Hook, NY,
  USA}, \bibinfo{pages}{43571--43608}.
\newblock
\urldef\tempurl%
\url{https://proceedings.neurips.cc/paper_files/paper/2024/hash/4ced59d480e07d290b6f29fc8798f195-Abstract-Conference.html}
\showURL{%
\tempurl}


\bibitem[Yuan et~al\mbox{.}(2023)]%
        {yuan2023morec}
\bibfield{author}{\bibinfo{person}{Zheng Yuan}, \bibinfo{person}{Fajie Yuan},
  \bibinfo{person}{Yu Song}, \bibinfo{person}{Youhua Li},
  \bibinfo{person}{Junchen Fu}, \bibinfo{person}{Fei Yang},
  \bibinfo{person}{Yunzhu Pan}, {and} \bibinfo{person}{Yongxin Ni}.}
  \bibinfo{year}{2023}\natexlab{}.
\newblock \showarticletitle{Where to Go Next for Recommender Systems? {ID}- vs.
  Modality-based Recommender Models Revisited}. In
  \bibinfo{booktitle}{\emph{Proceedings of the 46th International ACM SIGIR
  Conference on Research and Development in Information Retrieval}}.
  \bibinfo{publisher}{Association for Computing Machinery},
  \bibinfo{address}{New York, NY, USA}, \bibinfo{pages}{2639--2649}.
\newblock
\href{https://doi.org/10.1145/3539618.3591932}{doi:\nolinkurl{10.1145/3539618.3591932}}


\bibitem[Zeng et~al\mbox{.}(2026)]%
        {zeng2026agenticdata}
\bibfield{author}{\bibinfo{person}{Xingshan Zeng}, \bibinfo{person}{Zishan Xu},
  \bibinfo{person}{Boju Zhang}, \bibinfo{person}{Yuzhou Wu},
  \bibinfo{person}{Lingzhi Wang}, \bibinfo{person}{Jianghao Lin},
  \bibinfo{person}{Liangyou Li}, \bibinfo{person}{Yasheng Wang},
  \bibinfo{person}{Lifeng Shang}, \bibinfo{person}{Xin Jiang},
  \bibinfo{person}{Weinan Zhang}, \bibinfo{person}{Yong Yu},
  \bibinfo{person}{Qun Liu}, {and} \bibinfo{person}{Weiwen Liu}.}
  \bibinfo{year}{2026}\natexlab{}.
\newblock \bibinfo{title}{What Makes Good Agentic Data? An {ACE} Lens on Data
  Generation for {LLM} Agents}.
\newblock
\showeprint[arxiv]{2608.27260}
\href{https://doi.org/10.48550/arXiv.2608.27260}{doi:\nolinkurl{10.48550/arXiv.2608.27260}}


\bibitem[Zhang et~al\mbox{.}(2025)]%
        {zhang2025dgm}
\bibfield{author}{\bibinfo{person}{Jenny Zhang}, \bibinfo{person}{Shengran Hu},
  \bibinfo{person}{Cong Lu}, \bibinfo{person}{Robert Lange}, {and}
  \bibinfo{person}{Jeff Clune}.} \bibinfo{year}{2025}\natexlab{}.
\newblock \bibinfo{title}{Darwin G{\"o}del Machine: Open-Ended Evolution of
  Self-Improving Agents}.
\newblock
\showeprint[arxiv]{2505.22954}
\href{https://doi.org/10.48550/arXiv.2505.22954}{doi:\nolinkurl{10.48550/arXiv.2505.22954}}


\bibitem[Zhao et~al\mbox{.}(2021)]%
        {zhao2021recbole}
\bibfield{author}{\bibinfo{person}{Wayne~Xin Zhao}, \bibinfo{person}{Shanlei
  Mu}, \bibinfo{person}{Yupeng Hou}, \bibinfo{person}{Zihan Lin},
  \bibinfo{person}{Yushuo Chen}, \bibinfo{person}{Xingyu Pan},
  \bibinfo{person}{Kaiyuan Li}, \bibinfo{person}{Yujie Lu},
  \bibinfo{person}{Hui Wang}, \bibinfo{person}{Changxin Tian},
  \bibinfo{person}{Yingqian Min}, \bibinfo{person}{Zhichao Feng},
  \bibinfo{person}{Xinyan Fan}, \bibinfo{person}{Xu Chen},
  \bibinfo{person}{Pengfei Wang}, \bibinfo{person}{Wendi Ji},
  \bibinfo{person}{Yaliang Li}, \bibinfo{person}{Xiaoling Wang}, {and}
  \bibinfo{person}{Ji-Rong Wen}.} \bibinfo{year}{2021}\natexlab{}.
\newblock \showarticletitle{{RecBole}: Towards a Unified, Comprehensive and
  Efficient Framework for Recommendation Algorithms}. In
  \bibinfo{booktitle}{\emph{Proceedings of the 30th ACM International
  Conference on Information and Knowledge Management}}.
  \bibinfo{publisher}{Association for Computing Machinery},
  \bibinfo{address}{New York, NY, USA}, \bibinfo{pages}{4653--4664}.
\newblock
\href{https://doi.org/10.1145/3459637.3482016}{doi:\nolinkurl{10.1145/3459637.3482016}}


\bibitem[Zhou et~al\mbox{.}(2026)]%
        {zhou2026externalization}
\bibfield{author}{\bibinfo{person}{Chenyu Zhou}, \bibinfo{person}{Huacan Chai},
  \bibinfo{person}{Wenteng Chen}, \bibinfo{person}{Zihan Guo},
  \bibinfo{person}{Rong Shan}, \bibinfo{person}{Yuanyi Song},
  \bibinfo{person}{Tianyi Xu}, \bibinfo{person}{Yingxuan Yang},
  \bibinfo{person}{Aofan Yu}, \bibinfo{person}{Weiming Zhang},
  \bibinfo{person}{Congming Zheng}, \bibinfo{person}{Jiachen Zhu},
  \bibinfo{person}{Zeyu Zheng}, \bibinfo{person}{Zhuosheng Zhang},
  \bibinfo{person}{Xingyu Lou}, \bibinfo{person}{Changwang Zhang},
  \bibinfo{person}{Zhihui Fu}, \bibinfo{person}{Jun Wang},
  \bibinfo{person}{Weiwen Liu}, \bibinfo{person}{Jianghao Lin}, {and}
  \bibinfo{person}{Weinan Zhang}.} \bibinfo{year}{2026}\natexlab{}.
\newblock \bibinfo{title}{Externalization in {LLM} Agents: A Unified Review of
  Memory, Skills, Protocols and Harness Engineering}.
\newblock
\showeprint[arxiv]{2604.08224}
\href{https://doi.org/10.48550/arXiv.2604.08224}{doi:\nolinkurl{10.48550/arXiv.2604.08224}}


\end{thebibliography}

\clearpage
\appendix
\setlength{\intextsep}{6pt plus 2pt minus 2pt}
\section{Evaluation and Aggregation}
\label{app:evaluation}

\paragraph{Evaluation scope.}
Table~\ref{tab:overall} evaluates discovered algorithms in 16 settings from four recommenders and four datasets, combining task-specific search winners and source-selected designs retrained on target data. This assesses final quality and design reuse across environments. The common search cohort in Appendix~\ref{app:resources} examines progress and resources; source-linked bootstrap analysis below assesses search variability.

\paragraph{Final test performance and variability.}
For system $m$, evaluation setting $t$, and source-search repeat $r$, $\bar q_{m,t,r}$ averages final TEST retraining scores of the algorithm selected by development NDCG@5. Its design stays fixed during target retraining. Tables~\ref{tab:overall} and~\ref{tab:appendix-variation} report means and sample SDs over three searches:
\begin{equation}
\bar q_{m,t}=\frac{1}{3}\sum_{r=1}^{3}\bar q_{m,t,r},\qquad
s_{m,t}=\sqrt{\frac{1}{2}\sum_{r=1}^{3}(\bar q_{m,t,r}-\bar q_{m,t})^2}.
\end{equation}
The SD measures variation between discovered algorithms. OpenEvolve's zero SD on BERT4Rec/CDs, Electronics, and Office reflects three Seed-equivalent implementations sharing final test evaluations.

\begin{table*}[!t]
\centering
\caption{Search-to-search standard deviation of final TEST NDCG@5 and HR@5, aligned with Table~\ref{tab:overall}.}
\label{tab:appendix-variation}
\small
\renewcommand{\arraystretch}{0.88}
\setlength{\tabcolsep}{4pt}
\resizebox{\textwidth}{!}{%
\begin{tabular}{llcccccccc}
\toprule
\multirow{2}{*}{Seed Recommender} & \multirow{2}{*}{Method} & \multicolumn{2}{c}{CDs} & \multicolumn{2}{c}{Electronics} & \multicolumn{2}{c}{Office} & \multicolumn{2}{c}{MovieLens-1M} \\
\cmidrule(lr){3-4} \cmidrule(lr){5-6} \cmidrule(lr){7-8} \cmidrule(lr){9-10}
 &  & NDCG@5 & HR@5 & NDCG@5 & HR@5 & NDCG@5 & HR@5 & NDCG@5 & HR@5 \\
\midrule
\multirow{6}{*}{SASRec} & RecHarness & 0.0051 & 0.0114 & 0.0288 & 0.0335 & 0.0034 & 0.0047 & 0.0045 & 0.0039 \\
 & DeepEvolve & 0.0286 & 0.0356 & 0.0012 & 0.0019 & 0.0082 & 0.0154 & 0.0571 & 0.0538 \\
 & OpenEvolve & 0.0065 & 0.0074 & 0.0020 & 0.0035 & 0.0019 & 0.0011 & 0.0015 & 0.0034 \\
 & ShinkaEvolve & 0.0081 & 0.0142 & 0.0064 & 0.0057 & 0.0039 & 0.0084 & 0.0038 & 0.0052 \\
 & Self-EvolveRec & 0.0047 & 0.0105 & 0.0004 & 0.0010 & 0.0050 & 0.0087 & 0.0048 & 0.0048 \\
 & RecRSI (Ours) & 0.0095 & 0.0115 & 0.0185 & 0.0182 & 0.0032 & 0.0086 & 0.0049 & 0.0060 \\
\midrule
\multirow{6}{*}{MoRec} & RecHarness & 0.0156 & 0.0159 & 0.0051 & 0.0055 & 0.0055 & 0.0053 & 0.0021 & 0.0025 \\
 & DeepEvolve & 0.0149 & 0.0194 & 0.0097 & 0.0103 & 0.0022 & 0.0002 & 0.0077 & 0.0039 \\
 & OpenEvolve & 0.0026 & 0.0070 & 0.0003 & 0.0007 & 0.0019 & 0.0041 & 0.0015 & 0.0025 \\
 & ShinkaEvolve & 0.0077 & 0.0124 & 0.0073 & 0.0075 & 0.0033 & 0.0049 & 0.0018 & 0.0022 \\
 & Self-EvolveRec & 0.0072 & 0.0109 & 0.0033 & 0.0030 & 0.0019 & 0.0016 & 0.0016 & 0.0007 \\
 & RecRSI (Ours) & 0.0131 & 0.0170 & 0.0135 & 0.0155 & 0.0048 & 0.0048 & 0.0085 & 0.0057 \\
\midrule
\multirow{6}{*}{BERT4Rec} & RecHarness & 0.0108 & 0.0104 & 0.0072 & 0.0074 & 0.0122 & 0.0151 & 0.0061 & 0.0057 \\
 & DeepEvolve & 0.0082 & 0.0109 & 0.0041 & 0.0032 & 0.0086 & 0.0120 & 0.0091 & 0.0092 \\
 & OpenEvolve & 0.0000 & 0.0000 & 0.0000 & 0.0000 & 0.0000 & 0.0000 & 0.0068 & 0.0069 \\
 & ShinkaEvolve & 0.0006 & 0.0010 & 0.0085 & 0.0085 & 0.0049 & 0.0054 & 0.0153 & 0.0108 \\
 & Self-EvolveRec & 0.0408 & 0.0507 & 0.0118 & 0.0126 & 0.0311 & 0.0416 & 0.0103 & 0.0102 \\
 & RecRSI (Ours) & 0.0379 & 0.0389 & 0.0074 & 0.0083 & 0.0031 & 0.0064 & 0.0087 & 0.0059 \\
\midrule
\multirow{6}{*}{NGCF} & RecHarness & 0.0081 & 0.0118 & 0.0109 & 0.0124 & 0.0011 & 0.0025 & 0.0036 & 0.0116 \\
 & DeepEvolve & 0.0013 & 0.0035 & 0.0073 & 0.0121 & 0.0017 & 0.0024 & 0.0085 & 0.0192 \\
 & OpenEvolve & 0.0007 & 0.0014 & 0.0003 & 0.0003 & 0.0008 & 0.0015 & 0.0001 & 0.0004 \\
 & ShinkaEvolve & 0.0002 & 0.0001 & 0.0235 & 0.0280 & 0.0016 & 0.0020 & 0.0011 & 0.0028 \\
 & Self-EvolveRec & 0.0221 & 0.0221 & 0.0129 & 0.0163 & 0.0047 & 0.0035 & 0.0082 & 0.0134 \\
 & RecRSI (Ours) & 0.0006 & 0.0004 & 0.0205 & 0.0263 & 0.0075 & 0.0116 & 0.0001 & 0.0023 \\
\bottomrule
\end{tabular}%
}
\vspace{2pt}
\begin{minipage}{\textwidth}\footnotesize
Entries are sample SD ($n=3$) across source-search winners evaluated in the named environment. -- denotes an unavailable SD.
\end{minipage}
\end{table*}

\begin{table}[!t]
\centering
\small
\setlength{\tabcolsep}{3pt}
\caption{RecRSI versus each comparator on the 16 settings. $\Delta$ is mean relative gain (\%); W/T/L counts wins, ties, and losses based on mean test scores.}
\label{tab:head-to-head}
\begin{tabular*}{\columnwidth}{@{\extracolsep{\fill}}lrrrr@{}}
\toprule
& \multicolumn{2}{c}{NDCG@5} & \multicolumn{2}{c}{HR@5} \\
\cmidrule(lr){2-3}\cmidrule(lr){4-5}
Comparator & $\Delta$ & W/T/L & $\Delta$ & W/T/L \\
\midrule
RecHarness & +5.52 & 13/0/3 & +4.24 & 14/0/2 \\
DeepEvolve & +5.70 & 15/0/1 & +4.83 & 16/0/0 \\
OpenEvolve & +4.38 & 15/0/1 & +3.50 & 14/0/2 \\
ShinkaEvolve & +4.31 & 15/0/1 & +3.45 & 14/0/2 \\
Self-EvolveRec & +7.23 & 15/0/1 & +6.12 & 14/0/2 \\
\bottomrule
\end{tabular*}
\end{table}

\begin{table*}[!t]
\centering
\small
\setlength{\tabcolsep}{4pt}
\caption{Task data splits and ranking protocols. Sampled pools contain one positive and 99 fixed negatives; all candidates within a task share the same evaluation protocol.}
\label{tab:task-ranking}
\begin{tabular}{@{}p{0.25\textwidth}p{0.43\textwidth}p{0.27\textwidth}@{}}
\toprule
Task & Split / evaluation history & Ranking protocol \\
\midrule
SASRec / all four datasets & Per-user last-two holdout; test histories include training and validation interactions. & One positive plus 99 fixed negatives, evaluation seed 72026. \\
BERT4Rec/Electronics, Office & Amazon 2014 5-core; penultimate interaction for development, last for test. & Fixed uniform sampled pools. \\
BERT4Rec/MovieLens-1M & Per-user temporal split of the full MovieLens-1M data. & Fixed sampled pools; test negatives exclude train/dev and the test positive. \\
BERT4Rec/CDs & Common preprocessed data and user/item IDs across systems. & Fixed sampled candidate sets. \\
MoRec/CDs, Electronics & Common CDs preprocessing across systems; Electronics uses per-user last-two holdout. Test histories include training and validation interactions. & Fixed sampled candidate sets shared across systems. \\
MoRec/Office, MovieLens-1M & Per-user timestamp order (source-row tie-break); penultimate interaction for development, last for test. Test history: train + development. & One positive and 99 fixed uniform negatives; seed 72026. Negatives exclude the query history and target. \\
NGCF/CDs, Electronics, Office & Per-user last-two temporal holdout. & Fixed uniform pools; development/test seeds 72026/72027. \\
NGCF/MovieLens-1M & Fixed task split. & Full-catalog ranking. \\
\bottomrule
\end{tabular}
\end{table*}

\paragraph{Across-task comparisons.}
\label{app:head-to-head}
Relative gains are averaged equally over 16 settings, using Seed or the highest competing mean per setting and metric as the denominator. Table~\ref{tab:head-to-head} compares fixed systems. Against the strongest competitor, mean NDCG@5/HR@5 advantages remain positive when omitting any setting (1.96--2.94\%/1.66--2.56\%), starting model (1.57--3.35\%/1.35--2.97\%), or dataset (1.33--3.52\%/1.07--2.93\%).

\paragraph{Search-level robustness.}
Search variability is assessed on the eight MoRec and BERT4Rec settings with linked source-search and target-evaluation records. Mean NDCG@5/HR@5 gains over the task-wise strongest competitor are 3.70\%/3.06\%, with 95\% percentile bootstrap intervals [1.64\%, 5.57\%]/[1.36\%, 4.61\%]. In 100,000 draws, we independently resample searches within system and source task using aggregated retraining scores, carry each search's target evaluations together to preserve dependence, and recompute the strongest comparator. Both intervals exceed zero.

\section{Discovery Curves and Resource Accounting}
\label{app:resources}

\paragraph{Search progress and investment.}
Figures~\ref{fig:intro}(b) and~\ref{fig:discovery-cost} link progress and final quality to research expenditure using 144 trajectories: six systems, three searches, and eight tasks (SASRec/CDs; MoRec/CDs, Office, MovieLens-1M; BERT4Rec and NGCF each on Electronics and MovieLens-1M). Searches are averaged within tasks, then tasks equally. Figure~\ref{fig:intro}(b) tracks best development NDCG@5, including the initial algorithm and carrying the best forward after stopping. Figure~\ref{fig:discovery-cost}(c) plots its relative gain at cumulative input-plus-output token ceilings through 3 million, counting setup once and reported reasoning and cached tokens. Panels (a,b) pair final TEST gains with search costs; GPU efficiency divides mean relative gain by mean GPU-hours.

\paragraph{Budgets and stopping.}
The shared ceiling is 12 genuine evaluations; failed implementation or qualification consumes resources without counting as evaluation. A USD~7--8 API safeguard, fixed before search and independent of performance, stops searches at the cost limit using completed plus in-flight expenditure. It stopped 18 DeepEvolve or Self-EvolveRec searches (Table~\ref{tab:budget-evaluations}). Counts and costs include resumed work within each search; evaluation-budget exhaustion and execution interruptions also end searches.

\begin{table}[!htbp]
\centering
\small
\setlength{\tabcolsep}{3pt}
\caption{Completed evaluations in searches stopped by the API safeguard. R1--R3 are search repeats; -- marks a repeat not stopped by this safeguard.}
\label{tab:budget-evaluations}
\begin{tabular*}{\columnwidth}{@{\extracolsep{\fill}}llrrr@{}}
\toprule
System & Task & R1 & R2 & R3 \\
\midrule
\multirow{4}{*}{DeepEvolve} & MoRec/CDs & 2 & 6 & 0 \\
 & MoRec/Office & 0 & 1 & -- \\
 & MoRec/MovieLens-1M & 11 & -- & -- \\
 & BERT4Rec/Electronics & -- & 10 & 11 \\
\midrule
\multirow{5}{*}{Self-EvolveRec} & MoRec/CDs & 2 & -- & 10 \\
 & MoRec/MovieLens-1M & -- & -- & 1 \\
 & BERT4Rec/Electronics & 6 & 6 & -- \\
 & BERT4Rec/MovieLens-1M & 9 & 9 & -- \\
 & NGCF/Electronics & 7 & 0 & 2 \\
\bottomrule
\end{tabular*}
\end{table}

\paragraph{Search costs.}
Reference API cost includes research calls, initialization, and mechanism-space construction. GPU-hours cover allocation for qualification, each search's initial-training share, candidate fits, and unsuccessful attempts. Final TEST retraining and transfer evaluation are excluded from search cost.

\section{Mechanism Hypotheses and Feedback Controls}
\label{app:mechanism-records}

\paragraph{Mechanism space and feedback.}
Researchers fix source interfaces, editable targets, allowed axes, and composition limits. MechSpace generates primitives, parameter settings, and compatibility rules within them. For MoRec/CDs (Section~\ref{sec:method_mechspace}), generation proposes 18 primitives; review selects 16 across six axes. \texttt{SELECT} chooses one; \texttt{COMPOSE} combines two or three, with at most one per axis and two mutation targets: 16 singles, 106 pairs, and 372 triples. Primitives specify locations, types, stages, parameters, implementation, and incompatibilities. Explicit selection enforces these specifications; Prompt supplies them as text.

\paragraph{Four-dataset ablation.}
Table~\ref{tab:components} averages $100(q_{\max}/q_0-1)$ equally over four datasets, including the initial model in $q_{\max}$. All configurations complete 12 evaluations except MovieLens-1M Prompt+Summary (11); comparing all MovieLens-1M configurations at evaluation~11 gives unchanged frontiers. With $g_F,g_P,g_S,g_{P+S}$ denoting Full, Prompt, Summary, and Prompt+Summary gains, $g_F-g_P-g_S+g_{P+S}=0.6064$ percentage points measures how feedback changes the benefit of explicit mechanism choices.

\paragraph{Feedback comparisons.}
\label{app:feedback-controls}
Raw, Summary, and Mechanism share histories, legal actions, selection, implementation, and evaluation. Raw passes records; Summary gives research advice; Mechanism connects interventions and outcomes to follow-up hypotheses. Summary and Mechanism receive identical facts and source diagnostics with \texttt{gpt-5.6-terra}, medium reasoning, and a 20,000-token output limit.
The ten comparisons use frozen checkpoints from real MoRec/CDs research runs. For example, States~1--2 come from rounds~2--3 of the Full component-study search, and State~3 from round~18 of a continued Mechanism Feedback run.

\subsection{BERT4Rec Mechanism Hypotheses and Components}
\label{app:bert-case}
RO3 builds on label smoothing, which raised development NDCG@5 from 0.4045 to 0.4101, and responds to a residual-head regression by testing a new matching geometry. With query $\mathbf q_j$, tied item embeddings $E$, and existing output bias $\mathbf b$, its shared training/prediction scorer is
\begin{equation}
\mathbf s_j=(\mathbf q_jW)E^\top+\mathbf b,\qquad W\in\mathbb R^{64\times64}.
\label{eq:appendix-bilinear}
\end{equation}
The projection has no added bias. For $p_j=\operatorname{softmax}(\mathbf s_j)$, target $y_j$, catalog size $N$, and valid positions $\mathcal V$,
\begin{equation}
\mathcal L=-\frac{1}{|\mathcal V|}\sum_{j\in\mathcal V}
\left[0.9\log p_{j,y_j}+\frac{0.1}{N}\sum_{i=1}^{N}\log p_{j,i}\right].
\label{eq:appendix-smoothing}
\end{equation}
Padding is excluded; encoder, embeddings, and original output mapping remain unchanged. The hypothesis predicts that learning the matching geometry will complement the smoothed objective.

\paragraph{Component comparison design.}
Each column in Table~\ref{tab:bert-algorithm-factors} compares scoring and objective changes individually and together in one discovered design. RO1 also varies layer aggregation, giving eight combinations. These comparisons test how an intervention's contribution depends on accompanying mechanisms.

\begin{table}[H]
\centering
\small
\setlength{\tabcolsep}{4pt}
\caption{Component comparisons for BERT4Rec/Electronics designs RO1--RO3. A: layer aggregation; S: scoring change; O: full-catalog label smoothing.}
\label{tab:bert-algorithm-factors}
\begin{tabular*}{\columnwidth}{@{\extracolsep{\fill}}lrrr@{}}
\toprule
Design & RO1 & RO2 & RO3 \\
\midrule
Initial & 0.4079 & 0.4047 & 0.4045 \\
S & 0.4158 & 0.4379 & 0.4170 \\
O & 0.4107 & 0.4073 & 0.4101 \\
S + O & 0.4185 & \textbf{0.4398} & \textbf{0.4220} \\
A & 0.4098 & -- & -- \\
A + S & 0.4144 & -- & -- \\
A + O & 0.4133 & -- & -- \\
A + S + O & \textbf{0.4199} & -- & -- \\
\bottomrule
\end{tabular*}
\vspace{2pt}
\begin{minipage}{\columnwidth}\footnotesize
S is bilinear projection for RO1/RO3 and cosine-temperature scoring for RO2. A applies only to RO1.
\end{minipage}
\end{table}

\paragraph{Component contributions.}
RO3's combination reaches 0.4220, exceeding projection alone by 0.0051 and smoothing alone by 0.0119. Its interaction, $q_{SO}-q_S-q_O+q_0=-0.0005$, is approximately additive.
RO1 combines two-layer averaging (A), bilinear scoring (S), and smoothing (O). Its 0.4199 exceeds all seven other configurations. With S fixed, A changes quality by $-0.0014$ without O but $+0.0015$ with O, yielding conditional interaction $q_{ASO}-q_{AS}-q_{SO}+q_S=0.0029$.
RO2's cosine scorer (temperature 0.2) reaches 0.4379; smoothing adds 0.0019, reaching 0.4398, with interaction $-0.0007$.
Component comparisons use fixed training seeds 51006, 51007, and 51008 for RO1, RO2, and RO3, respectively.

\section{Task and LLM Configurations}
\label{app:task-configurations}

\paragraph{Data and training.}
Electronics and Office use Amazon Reviews 2014 5-core subsets~\cite{mcauley2015styles}; Table~\ref{tab:task-ranking} specifies splits and ranking protocols. Baseline search capabilities are preserved except web access. Within each starting-model setting, all systems share training and checkpoint selection; outer algorithm selection uses development NDCG@5.
SASRec: embedding 50, two causal blocks/two heads, length 200, dropout 0.2; Adam $10^{-3}$, betas $(0.9,0.98)$, batch 256, 300 epochs, evaluation every five epochs, patience four. One uniform negative per positive excludes training history.

BERT4Rec: two layers/heads, hidden/feed-forward 64/256, length 50, masking/dropout 0.2, learning rate $10^{-3}$, weight decay zero, train/evaluation batch 128/64; 100 epochs, NDCG@5 checks every five epochs, patience four.
MoRec: batch 256, hidden 50, length 200, dropout 0.2, embedding regularization 0.005; Adam $10^{-3}$, betas $(0.9,0.98)$, 300 epochs. Checkpoints are selected by development HR@5 every five epochs with patience four.
NGCF: learning rate $10^{-3}$, NDCG@5 each epoch, patience 30; batch/epochs 128/300 for CDs and Office, 1024/300 for Electronics, 1024/500 for MovieLens-1M.

\paragraph{Hardware and LLMs.}
Training and evaluation use NVIDIA RTX~3080~Ti GPUs. All discovery systems use GPT-5.6 Terra/Luna. Table~\ref{tab:llm-requests} gives the role assignments, with T for Terra and L for Luna. Research calls use medium reasoning with temperature left unset. Output limits follow the task and role configurations.

\begin{table}[H]
\centering
\small
\setlength{\tabcolsep}{3pt}
\caption{Role-specific LLM assignments.}
\label{tab:llm-requests}
\begin{tabular}{@{}p{0.25\columnwidth}p{0.70\columnwidth}@{}}
\toprule
System & Request roles \\
\midrule
RecHarness & \raggedright T: controller, implementation; L: research, critic. \tabularnewline
DeepEvolve & \raggedright T: planner, developer, reflection, writer; L: reader, search, debugger. \tabularnewline
OpenEvolve & \raggedright Generation: T and L. \tabularnewline
ShinkaEvolve & \raggedright Query: T and L; embedding: \texttt{text-embedding-3-small}. \tabularnewline
Self-EvolveRec & \raggedright T: development/debugging; L: research, diagnosis research, feedback. \tabularnewline
RecRSI & \raggedright L: Composer, Refiner, Falsification; T: Frontier, implementation. \tabularnewline
\bottomrule
\end{tabular}
\end{table}

\section{Research History and Subsequent Candidates}
\label{app:history-value}
\emph{With history} and \emph{History reset} start from the same MoRec/CDs incumbent (NDCG@5 0.4181), with prior experiments available or cleared. Access to research history improves candidate quality across all three successive decisions (Table~\ref{tab:history-value}). At the final step, the history branch keeps the scorer and clips item-embedding gradients to norm 0.5; the reset branch replaces BCE with pairwise logistic loss.

\begin{table}[H]
\centering
\small
\setlength{\tabcolsep}{3pt}
\caption{Research history improves successive candidate quality on MoRec/CDs. Values are development NDCG@5.}
\label{tab:history-value}
\begin{tabular*}{\columnwidth}{@{\extracolsep{\fill}}lrrr@{}}
\toprule
Round & With history & History reset & Difference \\
\midrule
13 & \textbf{0.4021} & 0.3686 & $+0.0334$ \\
14 & \textbf{0.3669} & 0.3405 & $+0.0265$ \\
15 & \textbf{0.4152} & 0.3953 & $+0.0199$ \\
\midrule
Mean & \textbf{0.3947} & 0.3681 & $+0.0266$ \\
\bottomrule
\end{tabular*}
\end{table}

\end{document}